\documentclass[12pt]{article}

\usepackage{epsfig}
\input epsf

\title{Charm photoproduction at HERA: $k_T$-factorization versus experimental data}

\author{A.V.~Lipatov, N.P.~Zotov}

\begin{document}

\maketitle

\begin{center}

{\it D.V.~Skobeltsyn Institute of Nuclear Physics,\\ 
M.V. Lomonosov Moscow State University,
\\119992 Moscow, Russia\/}\\[3mm]

\end{center}

\vspace{1cm}

\begin{center}

{\bf Abstract }

\end{center}

We calculate the cross section of charm photoproduction
at HERA collider in the framework of the $k_T$-factorization 
QCD approach. Our analysis cover the inclusive charm production
as well as charm and associated jet production processes. 
Both photon-gluon and gluon-gluon fusion mechanisms
are taken into account. 
The unintegrated gluon densities in a proton and in a photon 
 obtained from the full CCFM, from unified BFKL-DGLAP evolution 
equations as well as from the Kimber-Martin-Ryskin prescription 
are used. 
Our theoretical results are compared with the recent experimental data 
taken by the H1 and ZEUS collaborations at HERA.
Special attention is put on the specific angular correlations 
which can provide unique information about non-collinear gluon 
evolution dynamics. 

\vspace{1cm}

\section{Introduction} \indent 

The charmed quark production in electron-proton collisions 
at HERA is a subject of intensive study from both theoretical and experimental 
points of view~[1--6]. The value of charm mass $m_c$ provides a hard
scale which allows perturbative QCD (pQCD) to be applied.
The production dynamics is governed by the photon-gluon $\gamma g \to c\bar c$
or gluon-gluon fusion $gg \to c\bar c$ (direct and resolved photon 
contributions, respectively) and 
therefore cross sections of such processes are sensitive to the 
gluon content of a proton and of a photon. Very recently, the H1 and ZEUS collaborations 
have presented important experimental data~[5, 6] on the charm photoproduction at HERA.
In~[6] the data sample five times larger than in previous analysis~[1] has been used.
Differential cross sections are determined for events with a
$D^{*\pm}$ meson (inclusive $D^{*\pm}$ production) and for 
events with a $D^{*\pm}$ meson
and one or two hadronic jets. In the ZEUS analysis~[5] the differential 
inclusive jet cross sections for events containing a $D^{*\pm}$ 
meson have been measured and  specific angular correlations in the
$D^{*\pm}$ and dijet associated photoproduction have been studied.
A comparison of these measurements with the next-to-leading 
(NLO) pQCD calculations shows~[5, 6] that NLO pQCD has some
marked problems in description of experimental data.
In particular, the significant differences are observed~[5] in the shape and in
the normalization of most of the distributions between
data and theoretical predictions.
On the other hand recently much progress has been made towards
a global understanding of the $k_T$-factorization~[7, 8] (or semihard~[9, 10])
approach by working out this picture for several heavy quark
and prompt photon production processes at HERA and Tevatron~[11--15].

The $k_T$-factorization approach is based on the familiar Balitsky-Fadin-Kuraev-Lipatov 
(BFKL)~[16] or Ciafaloni-Catani-Fiorani-Marchesini (CCFM)~[17] gluon evolution.
In this way, the large logarithmic terms proportional to $\ln 1/x$ are summed 
up to all orders of perturbation theory (in the leading logarithmic 
approximation). It is in contrast with the popular 
Dokshitzer-Gribov-Lipatov-Altarelli-Parizi (DGLAP)~[18] strategy where only large 
logarithmic terms proportional to $\ln \mu^2$ are taken into account. 
The basic dynamical quantity of the $k_T$-factorization approach is 
the so-called unintegrated (i.e. ${\mathbf k}_T$-dependent) gluon distribution 
${\cal A}(x,{\mathbf k}_T^2,\mu^2)$ which determines the probability to find a 
gluon carrying the longitudinal momentum fraction $x$ and the transverse momentum 
${\mathbf k}_T$ at the probing scale $\mu^2$. The unintegrated gluon distribution
can be obtained from the analytical or numerical solution of the BFKL or CCFM
evolution equations. Similar to DGLAP, to calculate the cross sections of any 
physical process the unintegrated gluon density ${\cal A}(x,{\mathbf k}_T^2,\mu^2)$ 
has to be convoluted~[7--10] with the relevant partonic cross section. 
But as the virtualities of the propagating gluons are no longer ordered, the 
partonic cross section has to be taken off mass shell (${\mathbf k}_T$-dependent). 
It is in clear contrast with the usual DGLAP scheme (so-called collinear factorization). 
Since gluons in the initial state are not on-shell and are characterized by virtual 
masses (proportional to their transverse momentum), it also assumes a modification 
of their polarization density matrix~[7, 8]. In particular, the polarization 
vector of a gluon is no longer purely transversal, but acquires an admixture of 
longitudinal and time-like components. Other important properties of the 
$k_T$-factorization formalism are the additional contribution to the cross 
sections due to the integration over the ${\mathbf k}_T^2$ region above $\mu^2$
and the broadening of the transverse momentum distributions due to extra 
transverse momentum of the colliding partons.

Concerning the theoretical treatment of charm production in the
framework of standard (collinear) QCD, two types of NLO calculations are 
available for comparison with the recent H1 and ZEUS experimental data~[5, 6]. 
The traditional massive charm approach~[19] or fixed-flavour-number scheme (FFNS) 
assumes that light quarks are the only active flavours in the structure 
functions of the proton and photon, so that charmed quarks are produced 
only in the hard process. This scheme should be reliable when
the transverse momentum $p_T$ of the charmed quarks is of similar
size compared to $m_c$ and breaks down for $p_T \gg m_c$. It is because of
presence of collinear singularities which having the form $\alpha_s \ln (p_T^2/m_c^2)$. 
In the massless or zero-mass variable-flavour-number scheme
(ZMVFNS)~[20] charmed quarks are treated as an additional active flavours 
(massless partons). This approach is applicable at high transverse momenta $p_T \gg m_c$.
To completeness, we should also mention the general-mass variable-flavour-number
scheme (GMVFNS)~[21] which combines the massless and the massive scheme.
Note that the massless charm calculations take into account charm excitation processes 
and thus predict a larger resolved component in comparison with the 
massive calculations. However, both massless and massive approaches underestimate~[6] the
measured cross section of the inclusive $D^{*\pm}$ photoproduction
in the intermediate transverse momentum $p_T$ and forward pseudo-rapidity 
$\eta$ regions. An agreement between the theoretical and experimental
results can only be achieved using some extreme parameter values.
In particular, in the NLO massive scheme 
a very low charmed quark mass $m_c = 1.2$~GeV was required~[19].
But even within this set of parameters, the shapes of the 
$D^{*\pm}$ transverse momentum and pseudo-rapidity distributions
cannot be said well reproduced. Some better agreement between the 
massless scheme and the measured $p_T$ (though not $\eta$) spectrum
was achieved using special assumptions on the $c \to D^*$ fragmentation.
The similar situation is observed in the case of semi-inclusive 
charm production: the data tends to agree with the upper bound 
of the NLO calculation~[5]. However, the difference between the 
results of NLO calculation observed  in shape of 
inclusive differential cross section as function of the pseudo-rapidity 
$\eta_{D^{*\pm}}$ 
is not seen in the semi-inclusive cross section as a function of 
$\eta^{\rm jet}$, and the shape of the
data is well described by the NLO QCD predictions. At the same time the 
experimentally obtained dijet angular correlations~[5] show a large deviation from
the massive NLO QCD predictions, specially for the resolved-enriched sample.
In general, it was concluded~[6] that for the precise description of the charm
photoproduction higher-order corrections or implementation of
additional parton showers in current NLO calculations are needed.

In our previous paper~[11] the ability of the
$k_T$-factorization approach to reproduce the recent experimental
data for the $D^{*\pm}$ and dijet associated 
photoproduction (including the specific angular correlations between the hadronic 
jets in final state) taken by the ZEUS collaboration has been investigated.
It was demonstrated~[11, 22, 23] that the leading-order off-shell matrix elements 
of the photon-gluon fusion $|\bar {\cal M}|^2(\gamma g^* \to c\bar c)$ 
combined with the non-collinear
evolution of gluon densities in a proton ${\cal A}(x,{\mathbf k}_T^2,\mu^2)$ 
effectively simulate contribution from the charmed
quark excitation subprocess $cg \to cg$. Next, some $c - \bar c$ correlations
in high energy $\gamma p$ scattering have been studied~[24] and comparisons to the
recently measured data of the FOCUS collaboration at Fermilab were made.
In particular, it was shown that the analysis of the kinematical correlations
of charmed quarks opens new possibilities for verifying models
of non-collinear gluon evolution. 

In the present paper we will 
analyse the recent H1 and ZEUS data~[5, 6] using the $k_T$-factorization
approach of QCD. Mostly we will concentrate on 
the $D^{*\pm}$ and single jet associated photoproduction. 
It is because these processes have not been studied yet in the framework 
of $k_T$-factorization and the comparison to the
experimental data~[6] were made in the framework of MC generator 
\textsc{Cascade}~[25] only.  
We investigate the different production rates 
(calculated in a number of different kinematical regions) 
and make a systematic comparison of our predictions to the recent 
H1 and ZEUS data. Special attention will be drawn to the specific 
angular correlations in associated $D^{*\pm}$ and jet production since
these correlations are sensitive to the transverse momentum of the partons 
incoming to the hard scattering process and therefore sensitive to the 
details of the non-collinear gluon evolution.
Additional motivation of this study is the fact that $k_T$-factorization 
approach automatically incorporates the main part of the standard 
(collinear) high-order corrections~[7--10]. Our consideration will be based on 
the leading-order off-shell matrix elements of the photon-gluon and gluon-gluon 
fusion subprocesses (the direct and resolved photon contributions, respectively) 
which have been calculated in our previous papers~[12, 26]. 
In the numerical calculations we will test the different sets of 
unintegrated gluon distributions in a proton and in a photon which are 
obtained from the full CCFM~[27], from the unified BFKL-DGLAP evolution 
equations~[28] 
and from the conventional (DGLAP-based) quark and gluon densities. In the 
last case we will use the so-called Kimber-Martin-Ryskin (KMR)~[29] approach.

The outline of our paper is following. In Section~2 we 
recall shortly the basic formulas of the $k_T$-factorization approach with a brief 
review of calculation steps. In Section~3 we present the numerical results
of our calculations and a discussion. Finally, in Section~4, we give
some conclusions.

\section{Theoretical framework} 
\subsection{Kinematics} \indent 

We start from the gluon-gluon fusion subprocess. 
Let $p_e$ and $p_p$ be the four-momenta of the initial electron and 
proton, $k_1$ and $k_2$ the four-momenta of the incoming off-shell gluons, 
and $p_c$ and $p_{\bar c}$ the four-momenta of the produced
charmed quarks. In our analysis below we will use the Sudakov 
decomposition, which has the following form:
$$
  p_c = \alpha_1 p_e + \beta_1 p_p + p_{c\, T},\quad p_{\bar c} = \alpha_2 p_e + \beta_2 p_p + p_{\bar c\, T},\atop
  k_1 = x_1 p_e + k_{1T},\quad k_2 = x_2 p_p + k_{2T}, \eqno(1)
$$

\noindent 
where $k_{1T}$, $k_{2T}$, $p_{c\, T}$ and $p_{\bar c\, T}$ are the
transverse four-momenta of the corresponding particles.
It is important that ${\mathbf k}_{1T}^2 = - k_{1T}^2 \neq 0$ and
${\mathbf k}_{2T}^2 = - k_{2T}^2 \neq 0$. If we make replacement $k_1 \to p_e$
and set $x_1 = 1$ and $k_{1T} = 0$, then we easily obtain more
simpler formulas corresponding to photon-gluon fusion subprocess.
In the $ep$ center-of-mass frame we can write
$$
  p_e = {\sqrt s}/2 (1,0,0,1),\quad p_p = {\sqrt s}/2 (1,0,0,-1), \eqno(2)
$$

\noindent
where $s = (p_e + p_p)^2$ is the total energy of the process under consideration
and we neglect the masses of the incoming particles. The Sudakov variables
are expressed as follows:
$$
  \displaystyle \alpha_1={m_{c\, T}\over {\sqrt s}}\exp(y_c),\quad \alpha_2={m_{\bar c\, T}\over {\sqrt s}}\exp(y_{\bar c}),\atop
  \displaystyle \beta_1={m_{c\, T}\over {\sqrt s}}\exp(-y_c),\quad \beta_2={m_{\bar c\, T}\over {\sqrt s}}\exp(-y_{\bar c}), \eqno(3)
$$

\noindent
where $m_{c\, T}$ and $m_{\bar c\, T}$ are the transverse masses 
of the produced quarks, and $y_c$ and $y_{\bar c}$ 
are their rapidities (in the $ep$ center-of-mass frame). 
From the conservation laws we can easily obtain the following conditions:
$$
  x_1 = \alpha_1 + \alpha_2,\quad x_2 = \beta_1 + \beta_2,\quad {\mathbf k}_{1T} + {\mathbf k}_{2T} = {\mathbf p}_{c\, T} + {\mathbf p}_{\bar c \, T}. \eqno(4)
$$

\noindent 
In order to be sensitive to higher-order effects and to distinguish between
direct-enriched and resolved-enriched regions the variable $x_\gamma^{\rm obs}$ 
is often used~[5] in the analysis of the data which contain the jets. 
This variable, which is the fraction of the photon momentum contributing to 
the production of two jets with highest transverse energies $E_T^{\rm jet}$, 
is experimentally defined as
$$
  x_\gamma^{\rm obs} = { E_T^{{\rm jet}_1} e^{-\eta^{{\rm jet}_1}} + E_T^{{\rm jet}_2} e^{-\eta^{{\rm jet}_2}} \over 2 y E_e}, \eqno (5)
$$

\noindent
where $y E_e$ is the initial photon energy and $\eta^{{\rm jet}_i}$ are
the pseudo-rapidities of these hardest jets. 
The pseudo-rapidities $\eta^{{\rm jet}_i}$ are defined as
$\eta^{{\rm jet}_i} = - \ln \tan (\theta^{{\rm jet}_i}/2)$, where 
$\theta^{{\rm jet}_i}$ are the polar angles of the jets with respect to the proton beam.
The selection of $x_\gamma^{\rm obs} > 0.75$ and $x_\gamma^{\rm obs} < 0.75$ 
yields samples enriched in direct and resolved photon processes, respectively.
Another interesting variables, namely $x_\gamma^{\rm obs}(D^*)$ and 
$\cos \theta^*$, are also often used~[5, 6]
in the analysis of the experimental data. The $x_\gamma^{\rm obs}(D^*)$ 
variable can be constructed in 
an analogous way to the traditional $x_\gamma^{\rm obs}$~[5]. Using the $D^{*\pm}$
meson and the untagged jet (i.e. jet which is not matched to a $D^{*\pm}$ meson) 
of highest $E_T^{\rm jet}$, the quantity $x_\gamma^{\rm obs}(D^*)$ is given by
$$
  x_\gamma^{\rm obs}(D^*) = { p_T e^{-\eta} + E_T^{\rm jet} e^{-\eta^{\rm jet}} \over 2 y E_e}, \eqno (5)
$$

\noindent
where $p_T$ and $\eta$ are the transverse momentum and pseudo-rapidity
of the produced $D^{*\pm}$ meson. The scattering angle $\theta^*$ is
defined as
$$
  \cos \theta^* = \tanh{ \eta - \eta^{\rm jet}\over 2}. \eqno (11)
$$

\noindent Studying the distribution of $\cos \theta^*$ also gives us the 
possibility to learn about the size of the contributions from 
different production mechanisms~[5]. Finally, the inelasticity
$z$, defined by $z = (p_p \cdot p)/(p_p \cdot q)$ with
$p$ and $q$ being the four-momenta of the final $D^{*\pm}$
meson and the exchanged photon, is a measure of the fraction
of photon energy transferred to the $D^{*\pm}$ meson in the proton
rest frame. This quantity is sensitive to both the production
mechanism and to the $c \to D^*$ fragmentation details~[6].

\subsection{Cross section for charm photoproduction} \indent 

The main formulas for the total and differential cross sections for charm 
production cross sections 
were obtained in our previous papers~[11, 12, 26]. Here we recall some of 
them. In general case, the cross section $\sigma$ 
according to $k_T$-factorization theorem can be written as a convolution
$$
  \sigma = \int d{\mathbf k}_{T}^2 \, \hat \sigma({\mathbf k}_{T}^2,\mu^2) {\cal A}(x,{\mathbf k}_{T}^2,\mu^2), \eqno(6)
$$

\noindent
where $\hat \sigma({\mathbf k}_{T}^2,\mu^2)$ is the cross section
corresponding to the relevant partonic subprocess under consideration and 
${\cal A}(x,{\mathbf k}_{T}^2,\mu^2)$ is the unintegrated gluon distribution. 
The direct photon  contribution to the differential cross section of $\gamma 
p \to c\bar c + X$ process is given by
$$
  { d\sigma^{\rm (dir)} (\gamma p \to c\bar c + X) \over dy_c\, d{\mathbf p}_{c\, T}^2 } =
  \int {|\bar {\cal M}|^2(\gamma g^* \to c\bar c)\over 16\pi (x_2 s)^2 (1 - \alpha_1)} {\cal A}(x_2,{\mathbf k}_{2T}^2,\mu^2) d{\mathbf k}_{2T}^2 {d\phi_2 \over 2\pi} {d\phi_c \over 2\pi}, \eqno (7)
$$

\noindent
where $|\bar {\cal M}|^2(\gamma g^* \to c\bar c)$ is the squared off-shell matrix element which 
depends on the transverse momentum ${\mathbf k}_{2T}^2$, $\phi_2$ and $\phi_c$ are the 
azimuthal angles of the initial virtual gluon and the produced quark, respectively.
The formula for the resolved photon contribution can be obtained by the similar way. But
one should keep in mind that convolution in (6) should be made also with the
unintegrated gluon distribution ${\cal A}_\gamma(x,{\mathbf k}_{T}^2,\mu^2)$ in
a photon. The final expression for the differential cross section has the form
$$
  \displaystyle { d\sigma^{\rm (res)} (\gamma p \to c\bar c + X) \over dy_c\, d{\mathbf p}_{c\, T}^2 } = \int {|\bar {\cal M}|^2(g^* g^* \to c\bar c)\over 16\pi (x_1 x_2 s)^2} \times \atop 
  \displaystyle \times {\cal A}_\gamma(x_1,{\mathbf k}_{1T}^2,\mu^2) {\cal A}(x_2,{\mathbf k}_{2T}^2,\mu^2) d{\mathbf k}_{1T}^2 d{\mathbf k}_{2T}^2 dy_{\bar c} {d\phi_1\over 2\pi} {d\phi_2\over 2\pi} {d\phi_c\over 2\pi}, \eqno(8)
$$

\noindent
where $\phi_1$ is the azimuthal angle of the initial virtual gluon having fraction
$x_1$ of a initial photon longitudinal momentum. It is important that 
the squared off-shell matrix 
element $|\bar {\cal M}|^2(g^* g^* \to c\bar c)$ depends on the both transverse momenta 
${\mathbf k}_{1 T}^2$ and ${\mathbf k}_{2 T}^2$. 
The analytic expressions for the $|\bar {\cal M}|^2 (\gamma g^* \to c\bar c)$
and $|\bar {\cal M}|^2 (g^* g^* \to c\bar c)$ have been evaluated in our 
previous papers~[12, 26]. Note that if we 
average (7) and (8) over ${\mathbf k}_{1 T}$ and ${\mathbf k}_{2 T}$ and 
take the limit ${\mathbf k}_{1 T}^2 \to 0$ and ${\mathbf k}_{2 T}^2 \to 0$,
then we obtain well-known formulas corresponding to the leading-order (LO) QCD calculations.

The recent experimental data~[5, 6] taken by the H1 and ZEUS collaborations
refer to the $D^{*\pm}$ photoproduction in $ep$ collisions, where the 
electron is scattered
at small angle and the mediating photon is almost real ($Q^2 \sim 0$).
Therefore the $\gamma p$ cross sections (7) and (8) need to be weighted with 
the photon flux in the electron:
$$
  d\sigma(ep \to c\bar c + X) = \int f_{\gamma/e}(y)dy\, d\sigma(\gamma p \to c\bar c + X), \eqno (9)
$$

\noindent
where $y$ is a fraction of the initial electron energy taken by the photon in the 
laboratory frame, and we use the Weizacker-Williams approximation for the 
bremsstrahlung photon distribution from an electron:
$$
  f_{\gamma/e}(y) = {\alpha_{em} \over 2\pi}\left({1 + (1 - y)^2\over y}\ln{Q^2_{\rm max}\over Q^2_{\rm min}} + 
  2m_e^2 y\left({1\over Q^2_{\rm max}} - {1\over Q^2_{\rm min}} \right)\right). \eqno (10)
$$

\noindent
Here $\alpha_{em}$ is Sommerfeld's fine structure constant, $m_e$ is the electron 
mass, $Q^2_{\rm min} = m_e^2y^2/(1 - y)^2$ and $Q^2_{\rm max} \sim 1\,{\rm GeV}^2$, 
which is a typical value for the recent photoproduction measurements at HERA.

The multidimensional integration in (7), (8) and (9) has been performed
by means of the Monte Carlo technique, using the routine 
\textsc{Vegas}~[30]. The full C$++$ code is available from the authors on 
request\footnote{lipatov@theory.sinp.msu.ru}. This code is
practically identical to that used in~[11, 12], with exception
that now we apply it to calculate inclusive and jet(s) associated
charm production in another kinematical region.

\section{Numerical results} \indent 

We now are in a position to present our numerical results. First we describe our
theoretical input and the kinematical conditions.
As it was mentioned above, the recent experimental data~[5, 6] 
on the charm photoproduction at HERA come 
from both H1 and ZEUS collaboration. The ZEUS measurements~[5] are performed 
in the following kinematical region: $130 < W < 280$~ GeV, $Q^2 < 1$~GeV$^2$, 
$|\eta^{\rm jet}| < 2.4$, 
$E_T^{\rm jet} > 6$~GeV, $p_T > 3$~GeV and $-1.5 < \eta < 1.5$.
These data have been taken at a proton energy of 920~GeV and 
an electron energy of 27.5~GeV, which corresponds to a $e p$ center-of-mass 
(c.m.)
energy of $\sqrt s = 318$~GeV. Here and in the following all kinematic 
quantities are given in the laboratory frame where positive OZ axis 
direction is given by the proton beam. The more recent 
H1 data~[6] refer to the kinematical region defined by
$171 < W < 256$~ GeV, $Q^2 < 10^{-2}$~GeV$^2$, 
$|\eta^{\rm jet}| < 1.5$, $E_T^{\rm jet} > 3$~GeV, $p_T > 2$~GeV and 
$-1.5 < \eta < 1.5$.

\subsection{Theoretical uncertainties} \indent 

There are several parameters which determined the overall 
normalization factor of the cross sections (7) and (8): the charm 
mass $m_c$, the factorization and normalisation scales $\mu_F$ and $\mu_R$ 
and the unintegrated gluon distributions in a 
proton ${\cal A}(x,{\mathbf k}_T^2,\mu^2)$ and in a 
photon ${\cal A}_\gamma(x,{\mathbf k}_T^2,\mu^2)$. 

Concerning the unintegrated gluon densities in a proton, we have tried 
three different sets of the unintegrated gluon densities in a proton, 
namely J2003~(set~1)~[27], KMS~[28] and KMR~[29]. All these distributions are 
widely discussed in the literature (see, for example, review~[31--33] for 
more 
information). Here we only shortly discuss their characteristic properties.
First, the J2003~(set~1) gluon density has been obtained~[27] 
from the numerical solution of the full CCFM equation. The input parameters 
were fitted to describe the proton structure function $F_2(x,Q^2)$.
Note that this density contain only
singular terms in the CCFM splitting function $P_{gg}(z)$.
The J2003~(set~1) distribution has been applied in the analysis of the forward jet 
production at HERA and charm and bottom production at Tevatron~[27] 
(in the framework of Monte-Carlo generator \textsc{Cascade}~[25]) 
and has been also used in our calculations~[11, 12, 26].

Another set (the KMS)~[28] was obtained from a unified 
BFKL-DGLAP description of early $F_2(x, Q^2)$ data and includes the so-called 
consistency constraint~[34]. The consistency constraint introduces a 
large correction to the LO BFKL equation. It was argued~[34] that 
about 70\% of the full NLO corrections to the BFKL exponent 
$\Delta$ are effectively included in this constraint. 
The KMS gluon density is successful in description of the 
beauty hadroproduction at Tevatron~[15] and photoproduction
at HERA~[12].

The last, third unintegrated gluon distribution ${\cal A}(x,{\mathbf k}_T^2,\mu^2)$
used here (the so-called KMR distribution) 
is the one which was originally proposed in~[29]. The KMR approach is the 
formalism 
to construct unintegrated gluon distribution from the known conventional parton
(quark and gluon) densities. It accounts for the angular-ordering (which comes from 
the coherence effects in gluon emission) as well as the main part of the 
collinear higher-order QCD 
corrections. The key observation here is that the $\mu$ dependence of the unintegrated 
parton distribution enters at the last step of the evolution, and therefore single
scale evolution equations (DGLAP or unified BFKL-DGLAP) can be used up to this step. 
Also it was shown~[29] that the unintegrated distributions obtained via 
unified 
BFKL-DGLAP evolution are rather similar to those based on the pure DGLAP equations.
It is because the condition of the angular ordering constraint is more 
important~[29] than including the BFKL effects. Based on this point, 
in the present paper we use much more simpler DGLAP equation up to 
the last evolution step in the case of application of the KMR procedure. In 
the numerical calculations we have used the 
standard GRV~(LO) parametrizations~[35] of the collinear quark and gluon densities. 
Note that the KMR unintegrated parton distributions in a proton were used, 
in particular, to describe the prompt photon photoproduction at HERA~[13, 36] 
and prompt photon hadroproduction Tevatron~[14, 37].

In the case of a real photon, we have tested two different sets of the
unintegrated gluon densities ${\cal A}_\gamma(x,{\mathbf k}_T^2,\mu^2)$.
First of them was obtained~[38] from the 
numerical solution of the full CCFM equation (which has been also
formulated for the photon). Here we will use this gluon density together 
with the J2003~(set~1) distribution when calculating the resolved photon 
contribution (8). Also in order to obtain the unintegrated 
gluon density in a photon we will apply the KMR 
method~[29] to the standard LO GRV parton distributions~[35].
In the numerical calculations we will use it together with the KMR distributions
in a proton. Note that both gluon densities 
${\cal A}_{\gamma}(x,{\mathbf k}_T^2,\mu^2)$
discussed here have been already applied in the analysis of the  
charm and beauty quark~[26, 39] and $J/\psi$ meson production~[26] 
in $\gamma \gamma$ collisions at LEP2.

We would like to point out that at present there is no the unintegrated gluon
distribution corresponding to the unified BFKL-DGLAP evolution in a photon. 
Therefore the resolved photon contribution (8) is not taken into account 
in the case of KMS gluon distribution. 

Significant theoretical uncertainties in our results are connected with the
choice of the factorization and renormalization scales. The first of them
is related to the evolution of the gluon distributions, the other is 
responsible for the strong coupling constant $\alpha_s(\mu^2_R)$.
As it often done for charm production, we choose the 
renormalization and factorization scales to be equal: 
$\mu_R = \mu_F = \mu = \xi \sqrt{m_c^2 + \langle {\mathbf p}_{T}^2 \rangle}$, 
where $\langle {\mathbf p}_{T}^2 \rangle$ is set to the average
 ${\mathbf p}_{T}^2$ of the charmed quark and antiquark.
 In order to investigate the 
scale dependence of our results we will vary the scale parameter
$\xi$ between $1/2$ and 2 about the default value $\xi = 1$.
Note that we use special choice $\mu^2 = {\mathbf k}_T^2$ in the case of KMS 
gluon, as it was originally proposed in~[27]. 
The fragmentation $c \to D^*$ is described by Peterson
fragmentation function~[40] with $\epsilon_c = 0.035$~[41]. The branching
ratio $f(c \to D^*)$ was set to the value measured by the OPAL
collaboration: $f(c \to D^*) = 0.235$~[42].
For completeness, we take the charmed quark mass 
$m_c = 1.4$~GeV and use LO formula for the coupling constant $\alpha_s(\mu^2)$ 
with $n_f = 4$ active quark flavours at $\Lambda_{\rm QCD} = 200$~MeV, such 
that $\alpha_s(M_Z^2) = 0.1232$.

\subsection{Inclusive $D^{*\pm}$ production}
\indent

The results of our calculations are shown in Figs.~1 --- 4 in 
comparison to the recent H1 experimental data~[6]. Instead of presenting
our theoretical predictions as continuous lines, we adopt here the 
binning pattern encoded in the experimental data. The solid, dashed and 
dash-dotted histograms correspond to the results obtained 
with the J2003~(set~1), KMR and KMS unintegrated gluon densities, 
respectively. We observe a reasonable good agreement 
between our predictions and the H1 experimental data~[6] in most of the bins.
However, there are some insignificant discrepancies.
So, the calculated transverse momentum distribution
falls less steeply at large $p_T$ than it is visible
in the data. The similar effect was pointed out also in~[6] where  
the massive NLO pQCD approach~[18] (\textsc{fmnr} program~[43]) and the 
Monte-Carlo generator \textsc{cascade}~[35] have been used.
Concerning the dependence of our predictions on the
 evolution scheme, we have found that 
the $D^{*\pm}$ transverse momentum
distribution is only little sensitive to the choice of
unintegrated gluon density: very similar 
predictions are obtained by all parametrizations (see Fig.~1). 
This is in agreement with results of the previous investigations~[23].
The same situation is observed in $W$ distribution
where all gluon densities under consideration predict a very
similar shapes.
In contrast, the calculated pseudo-rapidity and $z$ distributions strongly
depend on the unintegrated gluon density used.
Note that pseudo-rapidity distribution additionally has been
measured~[6] in the three bins of $p_T$ (see Fig.~2).
In all these kinematical regions KMR gluon tends to
overestimate the data in the rear  direction
and KMS  tends to underestimate the data in the forward one.
This is, in particular, due to the fact that 
gluon-gluon fusion (resolved photon contribution)
is not taken into account in the case of KMS gluon in 
our calculations.
At the same time the pseudo-rapidity distributions obtained
with the J2003~(set~1) and KMR unintegrated gluon 
densities (where gluon-gluon fusion contribution 
is included everywhere) agree well with the H1 data~[6] in the forward 
pseudo-rapidity region. This is in  agreement with the general 
expectations that resolved photon contributions are
important at $\eta > 0$.

We would like to point out also that gluon-gluon
fusion contributes significantly in the low $z$ region.
In order to illustrate this effect we have separately 
shown the contributions from the photon-gluon (dashed histogram) 
and gluon-gluon fusion (dash-dotted histogram) mechanisms 
within the kinematic range of the H1 measurement~[6] (see Fig.~3). The solid 
histogram represents the sum of both these contributions.
We have used here the KMR unintegrated gluon density.
It is clear that gluon-gluon fusion mechanism is important 
at low $z$ and should be taken into account 
in description of the experimental data. 

Next, the total inclusive $D^{*\pm}$ photoproduction cross section 
$\sigma(ep \to e' D^{*\pm} + X)$ has been measured~[6] and 
it was found to be equal to $6.45 \pm 0.46$~(stat.)~$\pm 0.69$~(sys.)~nb.
The results of our calculations supplemented 
with the different unintegrated gluon densities are collected in 
Table~1. The theoretical uncertainties (which are
given for the J2003~(set~1) and KMR distributions) 
are connected with variation on the scale $\mu^2$ as it was described above.
The predictions of \textsc{cascade}, \textsc{pythia} 
as well as NLO pQCD calculations (GMVFNS approach 
and \textsc{fmnr} program) 
are shown for comparison. 
The central values from massive NLO pQCD calculations (\textsc{fmnr} 
program) and from \textsc{cascade} are slightly lower than
the measured result, whereas those of \textsc{pythia}~[44] and
GMVFNS~[20] are higher.
One can see that our predictions are 
rather close to ones from the \textsc{cascade} and 
\textsc{fmnr} programs and reasonably agree with the H1 data 
within the theoretical and experimental uncertainties.
Also we estimate the individual contributions 
from the photon-gluon and gluon-gluon fusion to the 
total cross section in the $k_T$-factorization approach.
We have found it to be about 80\% and 20\%, respectively.
Additionally we investigate the scale dependence of 
our predictions (see Fig.~4).
In these plots the solid histograms were obtained by 
fixing both the factorization and normalization scales at the default 
value $\mu^2$, whereas upper and lower dashed histograms 
correspond to the scale variation as it was described above.
Here we have used the KMR unintegrated gluon density.
One can see that scale variation changes the normalization of 
predicted cross section by 20 -- 30\% approximately. 
Note that we have not varied the charmed 
quark mass and used the default value of $m_c$ in all
calculations.

\begin{table}
\begin{center}
\begin{tabular}{|l|c|}
\hline
  Source & $\sigma(ep \to e' D^{*\pm} + X)$~[nb] \\
\hline
  H1 measurement~[6] & $6.45~\pm~0.46$~(stat.)~$\pm~0.69$~(sys.) \\
  \textsc{cascade}~[25] & $5.38^{+0.54}_{-0.62}$ \\
  \textsc{pythia}~[45] & $8.9$ \\
  \textsc{fmnr}~[44] & $5.9^{+2.8}_{-1.3}$ \\
  GMVFNS~[20] & $8.2^{+5.3}_{-4.0}$ \\
  J2003 (set 1) & $4.92^{+1.15}_{-0.72}$ \\
  KMR & $6.57^{+1.80}_{-1.48}$ \\
  KMS & 5.10 \\
\hline
\end{tabular}
\end{center}
\caption{The total cross section of the inclusive $D^{*\pm}$ 
  photoproduction in electron-proton collisions at 
  $Q^2 < 10^{-2}$~GeV$^2$, $0.29 < y < 0.65$, $p_T > 2$~GeV and 
  $-1.5 < \eta < 1.5$.}
\end{table}

\subsection{Associated $D^{*\pm}$ and single jet production}
\indent

Now we demonstrate how $k_T$-factorization approach can be 
used to calculate the semi-inclusive charm photoproduction rates.
The basic photon-gluon or gluon-gluon fusion subprocesses 
under consideration give rise to two high-energy charmed quarks, 
which can further evolve into hadron jets.
In our calculations the produced quarks (with their known kinematical 
parameters) were taken to play the role of the final jets. 
These two quarks are accompanied by a number of gluons radiated 
in the course of the gluon evolution. As it has been noted in~[22], on
the average the gluon transverse momentum decreases from the hard interaction
block towards the proton. As an approximation, we assume that the gluon 
emitted in the last evolution step and having the four-momenta $k'$ 
compensates the whole transverse momentum of the gluon participating in the hard 
subprocess, i.e. ${\mathbf k'}_T = - {\mathbf k}_T$. All the other emitted gluons are 
collected together in the proton remnant, which is assumed to carry only a negligible 
transverse momentum compared to ${\mathbf k'}_{T}$. This gluon gives rise to a 
final hadron jet with $E_T^{\rm jet} = |{\mathbf k'}_{T}|$ in addition to the jet 
produced in the hard subprocess. From these three hadron jets we choose the one jet
(or two jets) carrying the largest transverse energy, and then compute the charm 
and associated jet(s) production rates.

In the recent analysis~[5, 6] performed by of the H1 and ZEUS collaborations 
jets are divided into two categories: jets of the first
category are associated with the $D^{*\pm}$ meson ($D^*$-tagged jet),
while jets of the second category are not matched to a 
$D^{*\pm}$ meson ($D^*$-untagged jet). The inclusive,
$D^*$-tagged and untagged jet cross sections have been
measured~[5] by the ZEUS collaboration whereas H1 data refer to 
the $D^*$ and untagged jet cross sections only~[6]. Note that
formulation of the ZMVFNS do not provide the possibility
to simultaneously determine kinematic variables related to
the $D^{*\pm}$ meson and the jet that includes it, so that
a comparison to the massless NLO pQCD calculations is only
available for the untagged jet cross sections~[45].
In the following we will systematically compare the
predictions from the $k_T$-factorization approach
to all published data on associated charm and single jet production
at HERA.

\subsubsection{$D^*$-tagged jets: $p_T^{\rm jet}$ and $\eta^{\rm jet}$ distributions}
\indent

The results of our calculations are shown in Fig.~5 in 
comparison to the ZEUS experimental data~[5]. Notations of all histograms
here are the same as in Fig.~1. 
The pseudo-rapidity $\eta^{\rm jet}$ distribution
additionally has been
measured in the three bins of $E_T^{\rm jet}$.
One can see that shape
of the data is well described by our predictions, although
the normalisation is underestimated (by a factor of 1.5).
However, this discrepancy is not dramatic, because
some reasonable variations in charm mass $m_c$, energy scale 
$\mu^2$ or $\Lambda_{\rm QCD}$ parameter (not shown in Figs.) 
can partially cover the visible disagreement.
It is interesting that difference 
between the theoretical predictions calculated with  
different unintegrated gluon
densities in a proton are somewhat less pronounced in comparison
to the inclusive $D^{*\pm}$ production case. The 
KMS gluon tends to predict a larger cross sections than
ones obtained with other gluon densities under consideration.
This fact is in  
contrast with the results presented in the previous section where
KMR gluon dominates. The possible explanation is connected with the
difference in the kinematical region (in which our predictions
as well as experimental data were presented). 
We would like to point out also that our central predictions are 
very similar to the massive NLO pQCD results~[18]. It was
claimed~[5] that normalisation of the ZEUS data for all
distributions is reasonable described by the upper
limit of these NLO pQCD calculations.

\subsubsection{$D^*$-untagged jets: $x_{\gamma}^{\rm obs}(D^*)$ distribution}
\indent

Now we turn to the $D^{*\pm}$ and untagged jet production.
First we discuss the very interesting subject of study which is connected 
with the individual contributions from the direct and resolved photon 
mechanisms. As it was already mentioned 
above, the $x_\gamma^{\rm obs}(D^*)$ variable (which corresponds at leading 
order to the fraction of the exchanged photon momentum in the hard 
scattering process) provides a tool to investigate the relative 
importance of these different contributions. In LO collinear approximation, 
direct photon 
events at parton level have $x_\gamma^{\rm obs}(D^*) \sim 1$, 
while the resolved photon events populate the low values of $x_\gamma^{\rm obs}(D^*)$.
The same situation is observed in a NLO calculations, because in the three
parton final state any of these partons are allowed to take any kinematically
accessible value. In the $k_T$-factorization formalism the hardest transverse 
momentum parton emission can be anywhere in the evolution chain, and does 
not need to be closest to the photon as required by the 
strong $\mu^2$ ordering in DGLAP. Thus, if hardest jet originates from the 
$c\bar c$ pair, then $x_\gamma^{\rm obs}(D^*)$ is close to unity, but if a 
gluon from the initial cascade form the hardest transverse momentum jet, 
then $x_\gamma^{\rm obs}(D^*) < 1$. This statement is clearly demonstrated in Fig.~6 
where separately shown the contributions from the photon-gluon (dashed histogram) 
and gluon-gluon fusion (dash-dotted histogram) subprocesses within the kinematic
range of the ZEUS measurement~[5]. 
The solid histogram represents the sum of both these contributions.
We have used here the KMR unintegrated gluon density for illustration.
In agreement with the expectation for direct photon processes, the peak 
at high values of the $x_\gamma^{\rm obs}(D^*)$ is observed.
However, one can see that off-shell photon-gluon fusion results 
also in substantial tail at small values of $x_\gamma^{\rm obs}(D^*)$. 
The existence of this plateau in the collinear approximation of QCD usually is 
attributted to the charmed quark excitation from resolved photon.
In the $k_T$-factorization approach such plateau indicates the fact that the gluon 
radiated from evolution cascade appears to be harder than charmed quarks 
(produced in hard parton interaction) in a significant fraction of events~[11, 22].
Therefore we can conclude that the $k_T$-factorization formalism 
effectively imitates the charm component of the photon~[11, 22].
However, the predicted tail at small $x_\gamma^{\rm obs}(D^*)$ values is 
strongly depends on the unintegrated gluon distributions used, as
it was demonstrated in~[11]. Note that the gluon-gluon fusion events 
(with a gluon coming from the photon) are distributed over the 
whole $x_\gamma^{\rm obs}(D^*)$ range. It is clear that 
these events play role at small values of $x_\gamma^{\rm obs}(D^*)$ only and 
that now the total contribution to the cross section from the 
gluon inside the resolved photon is too small to be useful\footnote{This is 
one of reasons why the values of our cross sections (see Table~1 and~2) 
obtained with the J2003~(set~1) unintegrated gluon distribution are not more 
the \textsc{Cascade} predictions. The main reasons of this distinction are the 
different factorization scales in our calculations and the \textsc{Cascade} ones
and also the account for the initial and final showers in the MC generator 
\textsc{Cascade}~[6].}. This fact is 
in contrast with the inclusive $D^{*\pm}$ production case and coincide with 
the results~[45] obtained in the massless NLO QCD approximation.

In Fig.~7 we confront the calculated $x_\gamma^{\rm obs}(D^*)$ distribution
with the ZEUS data~[5]. 
One can see that our results corresponding to different gluon 
densities do not agree well
with the ZEUS data. If this disagreement is true, then
the possible explanation of this fact is the following.
The calculated cross sections at low $x_\gamma^{\rm obs}(D^*)$ are 
defined by the average value of the gluon transverse momenta 
$\langle k_{T} \rangle$ which is generated in the
course of the non-colliner evolution. It is due to the fact that 
events when 
the gluon jet has the largest $p_T$ among the three hadron jets 
contribute only in this kinematical region.
So, this average gluon $\langle k_{T} \rangle$ which 
generated by the all three versions of the unintegrated gluon 
distributions under consideration possible could be too small to describe the 
ZEUS data. The similar situation was observed in~[11] where the
$D^{*\pm}$ and dijet associated photoproduction has been considered.
However, the $k_T$-factorization approach well describes the
experimental data with the cut on the dijet invariant mass $M > 18$~GeV~[11].
It demonstrates that this cut is essential for applicability
of the description of resolved photon contributions by 
non-collinear evolution in a proton only.
Therefore we can conclude that further theoretical and experimental investigations
are necessary in 
order to understand and adequatively describe the ZEUS experimental data in the low 
$x_\gamma^{\rm obs}(D^*)$ region.

\subsubsection{$D^*$-untagged jets: $p_T$, $\eta$, $p_T^{\rm jet}$ and $\eta^{\rm jet}$ distributions}
\indent

In Figs.~8 --- 10 we confront the differential cross section
of $ep \to D^{*\pm} + {\rm jet} + X$ in photoproduction as 
measured~[5, 6] by H1 and ZEUS collaborations with our predictions.
One can see in Fig.~8 that the differential cross section as a function of $\eta$ and 
$p_T^{\rm jet}$ measured by the H1 collaboration is  
well described (both in shape and in magnitude) by our calculations.
Also the predicted cross section as a function of $\eta - \eta^{\rm jet}$
(see Fig.~10) are in a good agreement with the experimental data.
Note that there is some difference in shape of the $\eta^{\rm jet}$ distribution 
between the data and our calculations.
Similar to inclusive $D^{*\pm}$ production case, all 
unintegrated gluon densities under consideration slightly
overestimate the $d\sigma/dp_T$ distribution at 
large $p_T$ (see Fig.~8). But in general the H1 data 
on the total and differential cross sections
are reasonable well reproduced by our calculations.
In Table~2 the estimations of total cross section 
obtained from our calculations as well as from 
\textsc{cascade}, \textsc{pythia}, massive and massless
NLO pQCD evaluations are listed and compared with the H1 data~[6].
The theoretical uncertainties (which are
given for the J2003~(set~1) and KMR distributions) 
are connected with variation on the scale $\mu^2$, as it was 
made earlier in the inclusive $D^{*\pm}$ production case.

Concerning the ZEUS measurement, the situation
is not clear, as it seen from Fig.~9.
One can see that transverse energy distribution
is underestimated in all $E_T^{\rm jet}$ bins.
The difference in normalisation between our
predictions and the ZEUS data is about 1.5 or even 2.
This fact is in clear contrast with comparison of our
theoretical results with the H1 measurements.
Also the shape of measured $\eta^{\rm jet}$ distributions
is very different from the calculated one. Particularly, the ZEUS data  
are overshoot our theoretical estimations in the forward 
direction. But a good agreement is achieved 
in the rear pseudo-rapidity region (namely, at $\eta^{\rm jet} < 0.5$).
The similar situation is obtained~[5] in the framework of both
massive and massless NLO pQCD approaches~[43, 45].

\begin{table}
\begin{center}
\begin{tabular}{|l|c|}
\hline
  Source & $\sigma(ep \to e' D^{*\pm} + {\rm jet} + X)$~[nb] \\
\hline
  H1 measurement~[6] & $3.01~\pm~0.29$~(stat.)~$\pm~0.33$~(sys.) \\
  \textsc{cascade}~[25] & $3.08^{+0.22}_{-0.28}$ \\
  \textsc{pythia}~[45] & $3.8$ \\
  \textsc{fmnr}~[44] & $2.65^{+0.78}_{-0.42}$ \\
  ZMVFNS~[46] & $3.05^{+0.62}_{-0.47}$ \\
  J2003 (set 1) & $2.47^{+0.60}_{-0.39}$ \\
  KMR & $3.61^{+0.84}_{-0.68}$ \\
  KMS & 3.11 \\
\hline
\end{tabular}
\end{center}
\caption{The total cross section of the associated $D^{*\pm}$ and 
  single jet photoproduction in electron-proton collisions at
  $Q^2 < 0.01$~GeV$^2$, $0.29 < y < 0.65$, $p_T > 2$~GeV,
  $-1.5 < \eta < 1.5$, $p_T^{\rm jet} > 3$~GeV and
  $-1.5 < \eta^{\rm jet} < 1.5$.}
\end{table}

Now we turn to some qualitative comparison between the
predictions of $k_T$-factorization approach and the ones
obtained in the framework of collinear NLO pQCD approximation.
It is well known that in the case of inclusive single hadron 
photoproduction the direct and resolved photon contributions
are accumulated in the backward and forward directions, 
respectively. Using the ZMVFNS scheme, it was shown~[45] 
that the pseudo-rapidity 
$\eta^{\rm jet}$ of $D^{*}$-untagged jet (but not 
pseudo-rapidity $\eta$ of the $D^{*\pm}$ meson)
can serve as a discriminator between these 
two contributions. It is because the shapes of
direct and resolved photon mechanism are very
different to each other. In Fig.~11 we split our
central predictions for $d\sigma/d\eta^{\rm jet}$
distribution calculated in the ZEUS kinematical region 
into their direct and resolved  
photon components using the $x_\gamma^{\rm obs}(D^*)$ variable
and show the results
as the solid and dashed histograms, respectively. 
We have used the KMR
unintegrated gluon densities in a proton and in a photon. 
Similar to massless NLO pQCD~[45], we obtain a strong different
behaviour of calculated cross sections at 
$x_\gamma^{\rm obs}(D^*) > 0.75$ and 
$x_\gamma^{\rm obs}(D^*) < 0.75$.
Moreover, we reproduce well the direct contribution
both in normalization and shape (see also Fig.~8 from Ref.~[45]). 
But predicted resolved photon component lie
below NLO pQCD one by a factor of about 2, although
agree with NLO pQCD results in a shape.
The observed difference in normalisation 
is strongly depends, of course, on the unintegrated gluon distribution used.
In particular, difference between our calculations and 
results presented in~[45] indicates again that 
average gluon $\langle k_{T} \rangle$ 
is too small and that further theoretical investigations of non-collinear 
gluon evolution are needed. However, at qualitative level,
the $k_T$-factorization formalism in a simplest way reproduces 
well the basic properties of the standard fixed-order calculations
for process under consideration. 

\subsubsection{$D^*$-untagged jets: $\cos \theta^*$ variable}
\indent

Next we turn to the different angular correlations in 
the charm production at HERA. First we discuss the 
$\cos \theta^*$ variable.
As it was already mentioned above, studying of distribution on 
$\cos \theta^*$ 
also give us the possibility
to learn about the size of the contribution from different production 
mechanisms. It is because this quantity is sensitive to the spin of 
the propagator in the hard subprocess~[5]. In direct photon processes 
$\gamma g \to c\bar c$ the propagator in the LO QCD diagrams is a quark, 
and the differential cross section rises slowly towards high $|\cos \theta^*|$ 
values, namely proportional to $(1 - |\cos \theta^*|)^{-1}$.
In resolved processes, the gluon propagator is allowed at LO and 
dominates over the quark propagator due to the stronger gluon-gluon coupling 
compared to the quark-gluon coupling. If most of the resolved photon events 
are produced as a result of charm from the photon, a gluon-exchange 
contribution in $cg \to cg$ subprocess should dominate. This results in a 
steep rise of the cross section towards high $|\cos \theta^*|$ values: 
$d\sigma/d\cos\theta^* \sim (1 - |\cos \theta^*|)^{-2}$.
A recent ZEUS analysis~[5] on dijet angular distributions in the $D^{*\pm}$
photoproduction has shown that the measured cross section from
resolved-enriched events (i.e. events with $x_\gamma^{\rm obs}(D^{*}) < 0.75$)
exhibits a distinct asymmetry with a strong rise towards $\cos \theta^* = - 1$.
This behaviour suggest that events with $x_\gamma^{\rm obs}(D^{*}) < 0.75$
are dominantly produced by charmed quarks coming from the 
photon side. On the other hand, the $\cos \theta^*$ distribution
for direct-enriched events (where $x_\gamma^{\rm obs}(D^{*}) > 0.75$) 
is almost symmetric. In the $k_T$-factorization approach 
the $\cos \theta^*$ distribution is 
determined only by the photon-gluon fusion off-shell matrix 
element which cover both scattering process (since there is no restriction on 
the transverse momenta along the evolution cascade, as it was already 
discussed above). In order to study this quantity in a detail,
in Fig.~12 we show the differential 
cross section as a function of $\cos \theta^*$ for the direct-enriched (solid 
histogram) and resolved-enriched (dashed histogram) samples separately 
within the kinematic range of the ZEUS experiment. We have used here the KMR 
unintegrated gluon density for illustration. One can see that 
resolved photon-like events exhibit a strong rise 
towards $\cos\theta^* = - 1$, i.e. in
photon direction, consistent with a dominant contribution from
gluon exchange. In our theoretical calculations, the peak at 
$\cos\theta^* = - 1$ at low $x_\gamma^{\rm obs}(D^{*})$ clearly illustrates
again that the $k_T$-factorization approach effectively reproduces the 
charm excitation processes using only the photon-gluon fusion 
off-mass shell matrix elements. However, we should point out that 
the absolute normalization of this peak differs from the one 
calculated~[45] in the massless NLO pQCD approximation (by a factor of about 2). 
This fact is in a full agreement with the results presented in the 
previous section (in Fig.~11) and indicates again that resolved photon 
contribution is underestimated by our calculations (compared to
the NLO pQCD ones). Unfortunately, at present there 
is no experimental data on $\cos \theta^*$ distribution in the 
associated $D^{*\pm}$ and single jet photoproduction. 
Therefore additional experimental efforts in this field
possible can give us the possibility to better constraint 
the unintegrated gluon density in a proton.

\subsubsection{$D^*$-untagged jets: azimuthal correlations}
\indent

Further understanding of the process dynamics and in particular of 
the high-order correction effects may be obtained from the transverse 
correlation between the produced $D^{*\pm}$ meson and the jet.
The H1 collaboration has measured~[6] the distribution on the 
$\Delta \phi$ angle, which is the difference in azimuth between 
the $D^{*\pm}$ meson and the jet.
In the naive leading order approximation, the distribution 
over $\Delta \phi$ must be simply a delta function 
$\delta(\Delta \phi - \pi)$, since the produced $D^{*\pm}$ meson 
and the jet are back-to-back in the transverse plane. Taking 
into account the non-vanishing initial parton
transverse momenta ${\mathbf k}_{1 T}$ and ${\mathbf k}_{2 T}$ leads to 
the violation of this back-to-back kinematics in the $k_T$-factorization 
approach~[15].

The calculated $\Delta \phi$ distributions are shown in Fig.~13
in comparison to the H1 data. Additionally, in Figs.~14 and~15 
we plot our predictions separately for the regions 
$x_\gamma^{\rm obs}(D^{*}) > 0.75$ and $x_\gamma^{\rm obs}(D^{*}) < 0.75$, 
where direct and resolved photon induced processes dominate, respectively.
Last results have been calculated for the ZEUS kinematical range,
as it was described above. 
It is clear that $k_T$-factorization approach predicts a large 
fraction of events where the $D^{*\pm}$ and the jet are 
not back-to-back. The measured deviations from the 
back-to-back topology are reasonable well described by 
our calculations. Best agreement is achieved using the KMR 
unintegrated gluon density. There is general behaviour
of theoretical results obtained using the unintegrated
gluon densities under consideration. All of them 
overestimate the H1 data at $\Delta \phi \sim 0$ (where 
gluon ${\mathbf k}_{T}$ is large) as well as 
at $\Delta \phi \sim \pi$, i.e. in the back-to-back region.
But at moderate $\Delta \phi$ our predictions
tends to underestimate the data. One can see
that region of low $\Delta \phi$ is very sensitive to
the unintegrated gluon distributions: the difference 
between predictions can be one order of magnitude (see Figs.~14 
and~15). The similar results has been obtained in~[11, 15]. 
Therefore we can expect that further theoretical and 
experimental studying of 
these correlations will give important information about 
non-collinear parton evolution
dynamics in a proton and in a photon.

\section{Conclusions} \indent 

We have investigated the charm photoproduction in electron-proton collisions 
at HERA using the $k_T$-factorization approach of QCD.
Our analysis cover the inclusive charm production
as well as charm and associated jet production processes. 
Both photon-gluon (direct) and gluon-gluon fusion (resolved) mechanisms
were taken into account. Our investigations were based on the 
leading-order off-shell matrix elements of the relevant partonic 
subprocesses. The total and differential cross sections have been 
calculated and the comparisons to the recent H1 and ZEUS experimental 
data have been made.
In numerical analysis we have used the unintegrated gluon densities 
which are obtained from the full CCFM, from unified BFKL-DGLAP 
evolution equations as well as from the Kimber-Martin-Ryskin 
prescription. 

We have shown that the $k_T$-factorization approach
reproduces reasonably well the numerous H1 data
on both inclusive charm and associated charm and jet production. 
At the same time we find that the ZEUS measurements 
overshoot our theoretical predictions by a factor of about 1.5 or even 2. 
It means that additional efforts should be done from both the theoretical
and experimental sides in order to reduce this
disagreement. 

We find that our results agree with the standard NLO pQCD ones
at qualitative level for process under consideration. It was demonstrated that 
off-shell matrix elements combined with the non-collinear 
evolution of gluon densities in a proton effectively simulate 
the charmed quark excitation processes. We find that 
resolved photon contributions are underestimated 
in the framework of $k_T$-factorization formalism compared
to the massless NLO pQCD calculations.
Of course, degree of this underestimation strongly depends
on the unintegrated gluon density used.

Special attention in our calculations
has been drawn to the specific 
angular correlations between the produced $D^{*\pm}$ meson and 
jets in final state. In particular, we demostrate the strong 
sensitivity of the $\Delta \phi$ distribution at low $\Delta \phi$ 
to the unintegrated gluon densities in a proton and in a photon.

To conclude, we believe that the $k_T$-factorization formalism 
holds a possible key to understanding charm production at HERA.
However, there are still large uncertainties, and much more
work needs to be done.

\section{Acknowledgements} \indent 
We thank H.~Jung for possibility to use the CCFM code for inintegrated dluon 
distributions in our calculations, for careful reading of the manuscript
and critical remarks.
The authors are very grateful to S.P.~Baranov 
for encouraging interest and very helpful discussions, S.~Chekanov and 
J.~Loizides for discussion of the ZEUS data,
P.F.~Ermolov for the support and DESY Directorate for the
support and the hospitality.
A.V.L. was supported in part by the grant of President of 
Russian Federation (MK-9820.2006.2). 
Also this research was supported by the 
FASI of Russian Federation (grant NS-8122.2006.2).

\newpage

\begin{figure}
\epsfig{figure=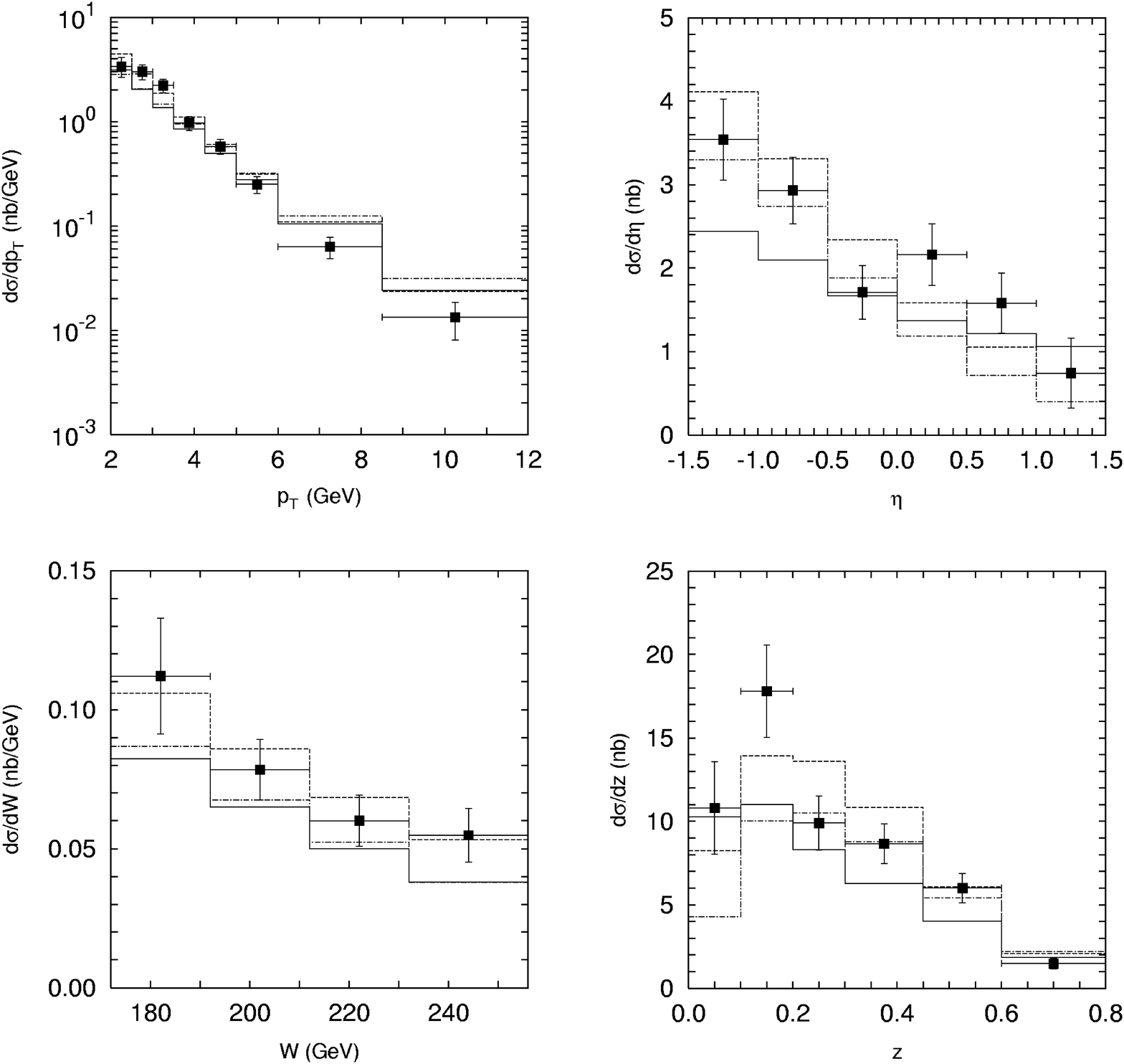, width = 17cm}
\caption{The differential cross sections
of the inclusive $D^{*\pm}$ production at HERA calculated 
in the kinematic range $Q^2 < 0.01$~GeV$^2$, $0.29 < y < 0.65$, 
$p_T > 2$~GeV and $-1.5 < \eta < 1.5$. The solid, 
dashed and dash-dotted histograms correspond to the 
predictions obtained with the J2003~(set~1), KMR and KMS unintegrated 
gluon densities, respectively. The experimental data are from H1~[6].}
\label{fig1}
\end{figure}

\newpage

\begin{figure}
\epsfig{figure=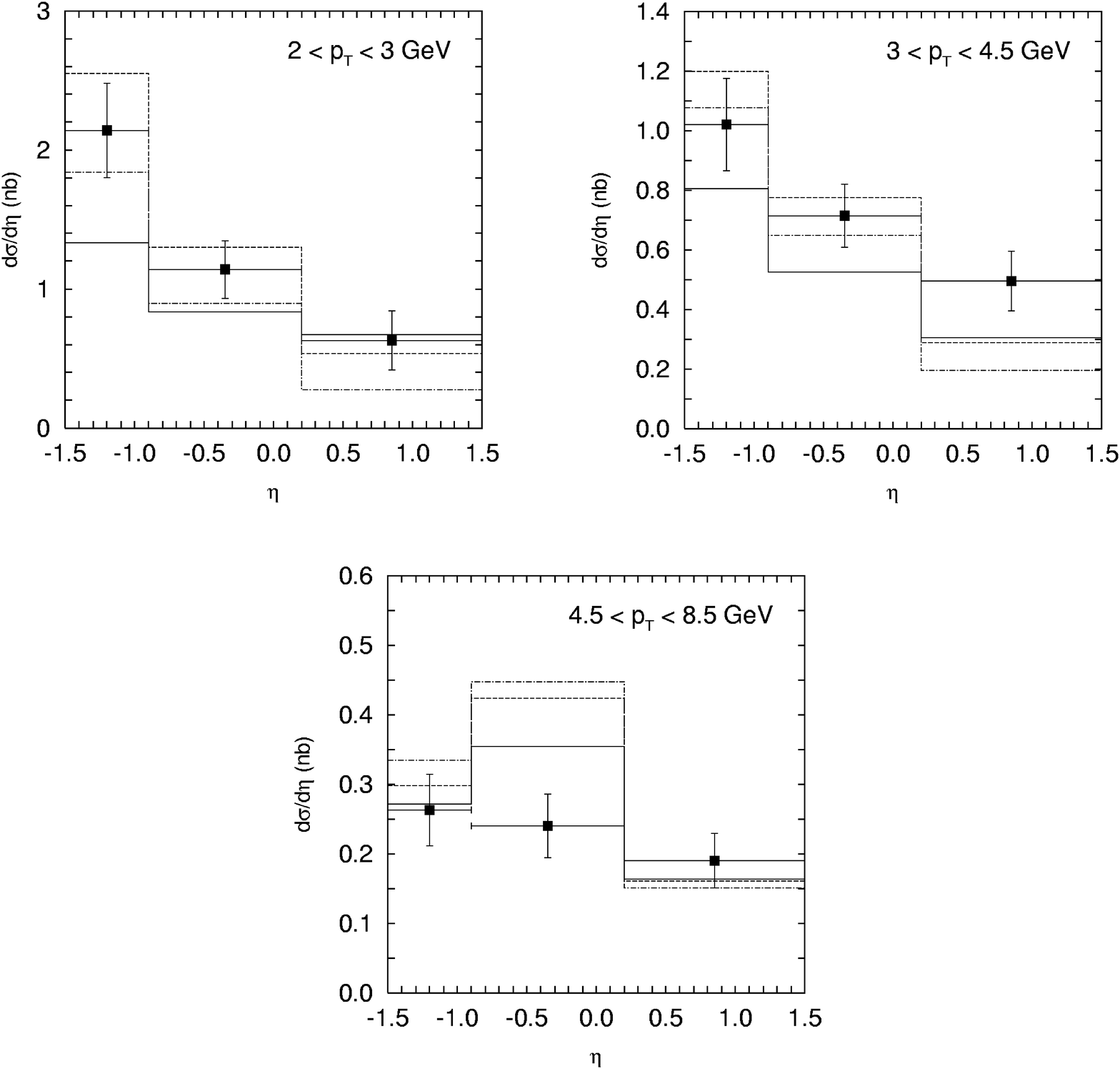, width = 17cm}
\caption{Inclusive $D^{*\pm}$ cross sections as a function 
of $\eta$ for three bins of $p_T$. Notations of all histograms here are the 
same as in Fig.~1. The experimental data are from H1~[6].}
\label{fig2}
\end{figure}

\newpage

\begin{figure}
\epsfig{figure=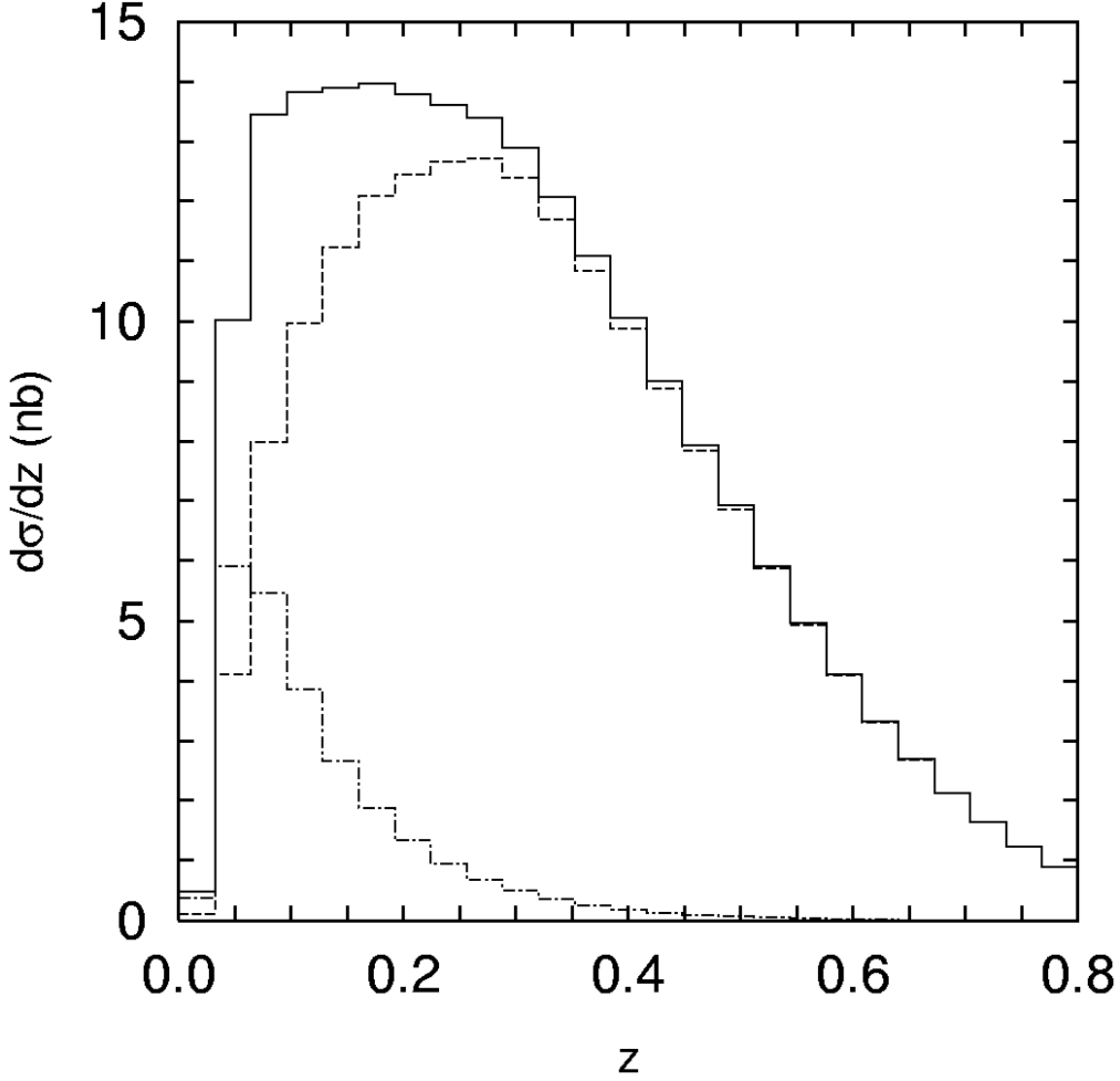, width = 22cm}
\caption{The $d\sigma/dz$ distribution of the 
inclusive $D^{*\pm}$ production at HERA calculated in the kinematic range 
$Q^2 < 0.01$~GeV$^2$, $0.29 < y < 0.65$, $p_T > 2$~GeV and 
$-1.5 < \eta < 1.5$. Separately shown the contributions
from the photon-gluon (dashed histogram) and gluon-gluon
(dash-dotted histogram). Solid histogram represents the
sum of both these contributions. The KMR unintegrated
gluon densities in a proton and in a photon have been used.}
\label{fig3}
\end{figure}

\newpage

\begin{figure}
\epsfig{figure=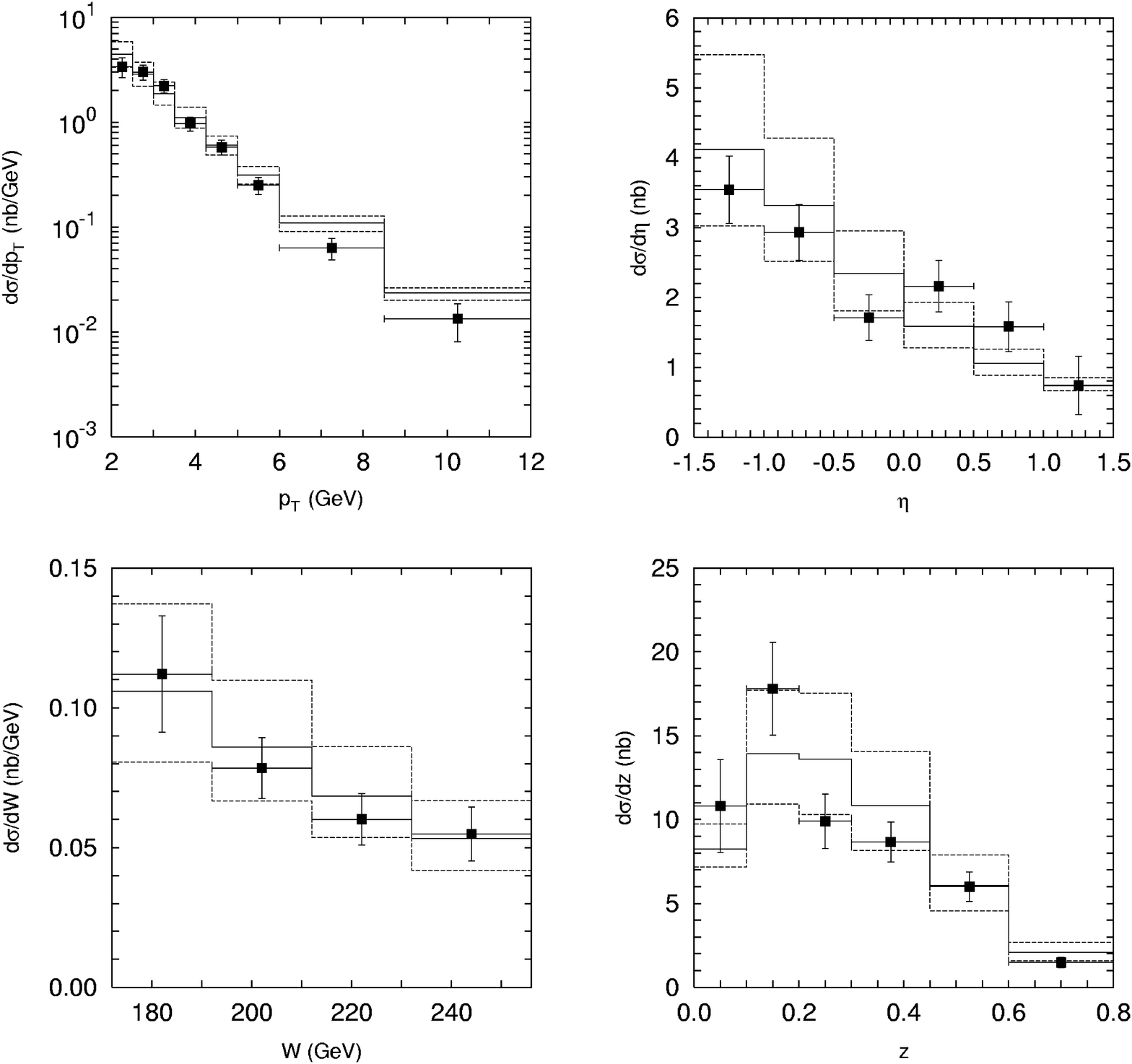, width = 17cm}
\caption{The $d\sigma/dp_T$ distribution of the 
inclusive $D^{*\pm}$ production at HERA calculated 
in the kinematic range 
$Q^2 < 0.01$~GeV$^2$, $0.29 < y < 0.65$ and 
$-1.5 < \eta < 1.5$. Solid, upper dashed and lower dashed histograms 
were obtained with the $\mu^2 = m_T^2$, $\mu^2 = 1/4 m_T^2$ and
$\mu^2 = 4 m_T^2$, respectively. The experimental data are from H1~[6].}
\label{fig4}
\end{figure}

\newpage

\begin{figure}
\epsfig{figure=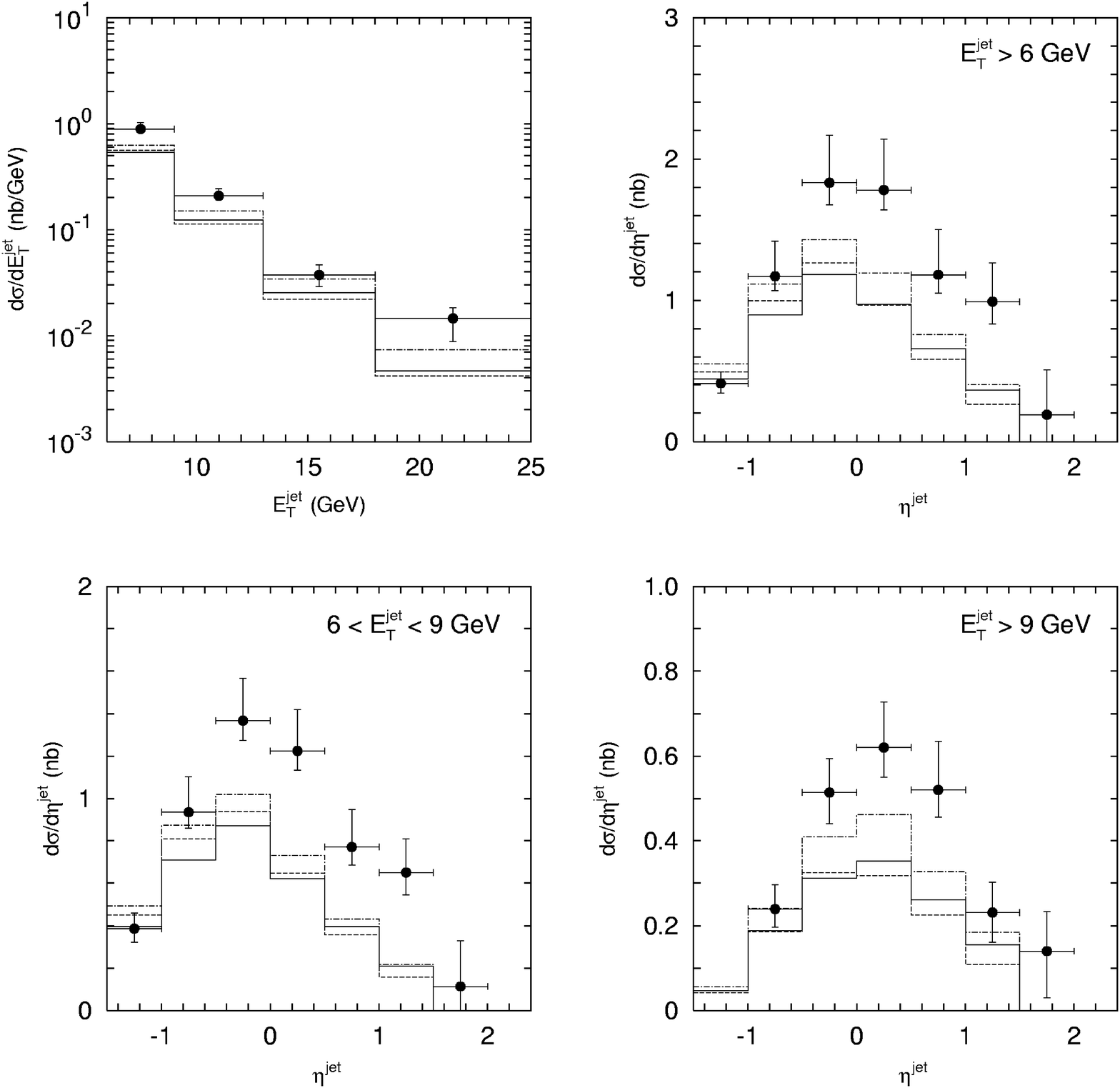, width = 17cm}
\caption{The differential cross sections
of the $D^{*\pm}$ and tagged jet production at 
HERA calculated in the kinematic range 
$Q^2 < 1$~GeV$^2$, $130 < W < 280$~GeV, 
$p_T > 3$~GeV, $|\eta| < 1.5$, $-1.5 < \eta^{\rm jet} < 2.4$
and $E_T^{\rm jet} > 6$~GeV. Notations of all histograms here are the 
same as in Fig.~1. The experimental data are from ZEUS~[5].}
\label{fig5}
\end{figure}

\newpage

\begin{figure}
\epsfig{figure=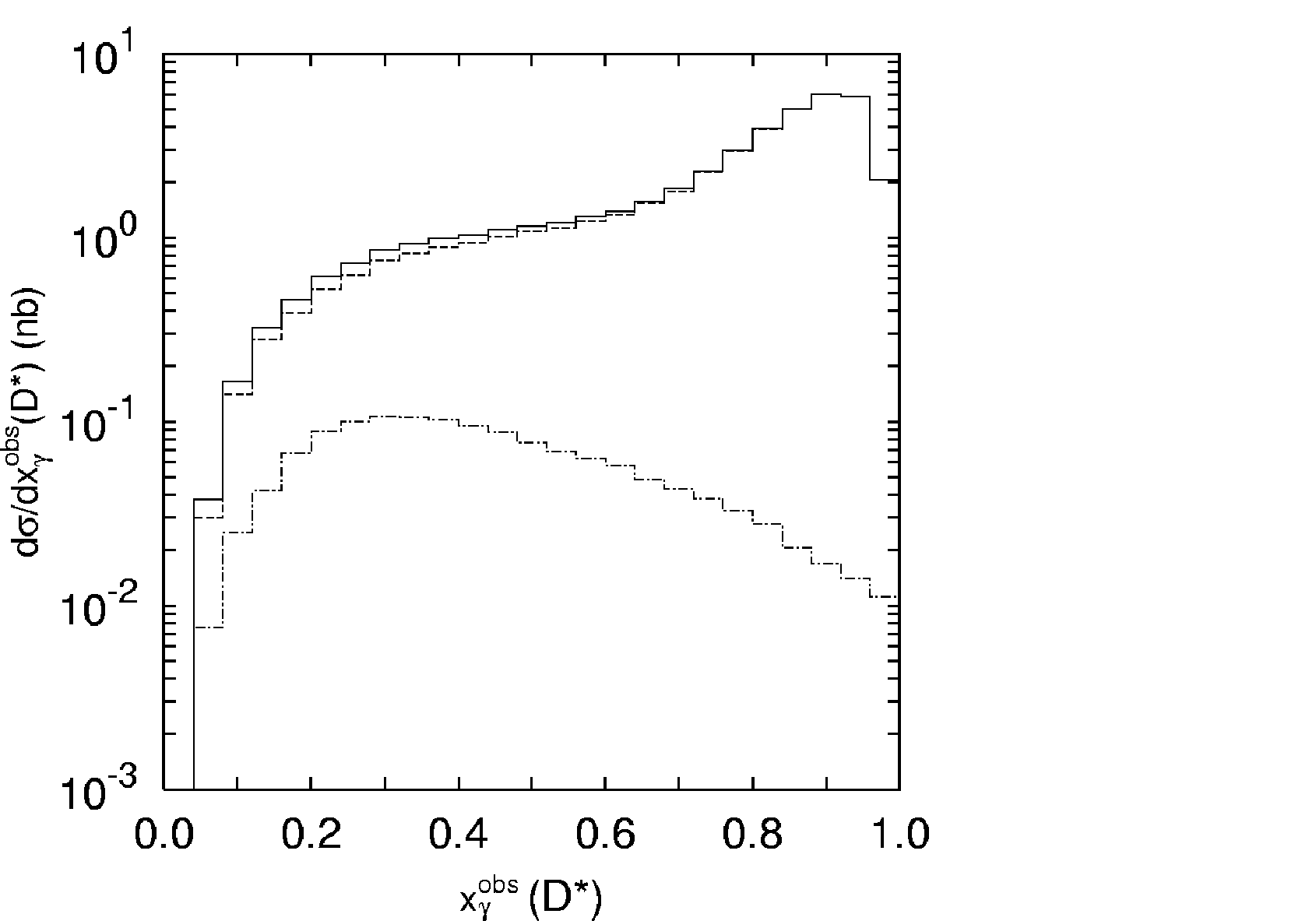, width = 22cm}
\caption{Direct (dashed histogram) and resolved photon 
(dashed-dotted histogram) components of our central predictions
of the differential cross section $d\sigma/dx_\gamma^{\rm obs}(D^*)$
calculated in the kinematic range 
$Q^2 < 1$~GeV$^2$, $130 < W < 280$~GeV, 
$p_T > 3$~GeV, $|\eta| < 1.5$, $-1.5 < \eta^{\rm jet} < 2.4$
and $E_T^{\rm jet} > 6$~GeV. The KMR unintegrated
gluon densities in a proton and in a photon have been used.}
\label{fig5a}
\end{figure}

\newpage

\begin{figure}
\epsfig{figure=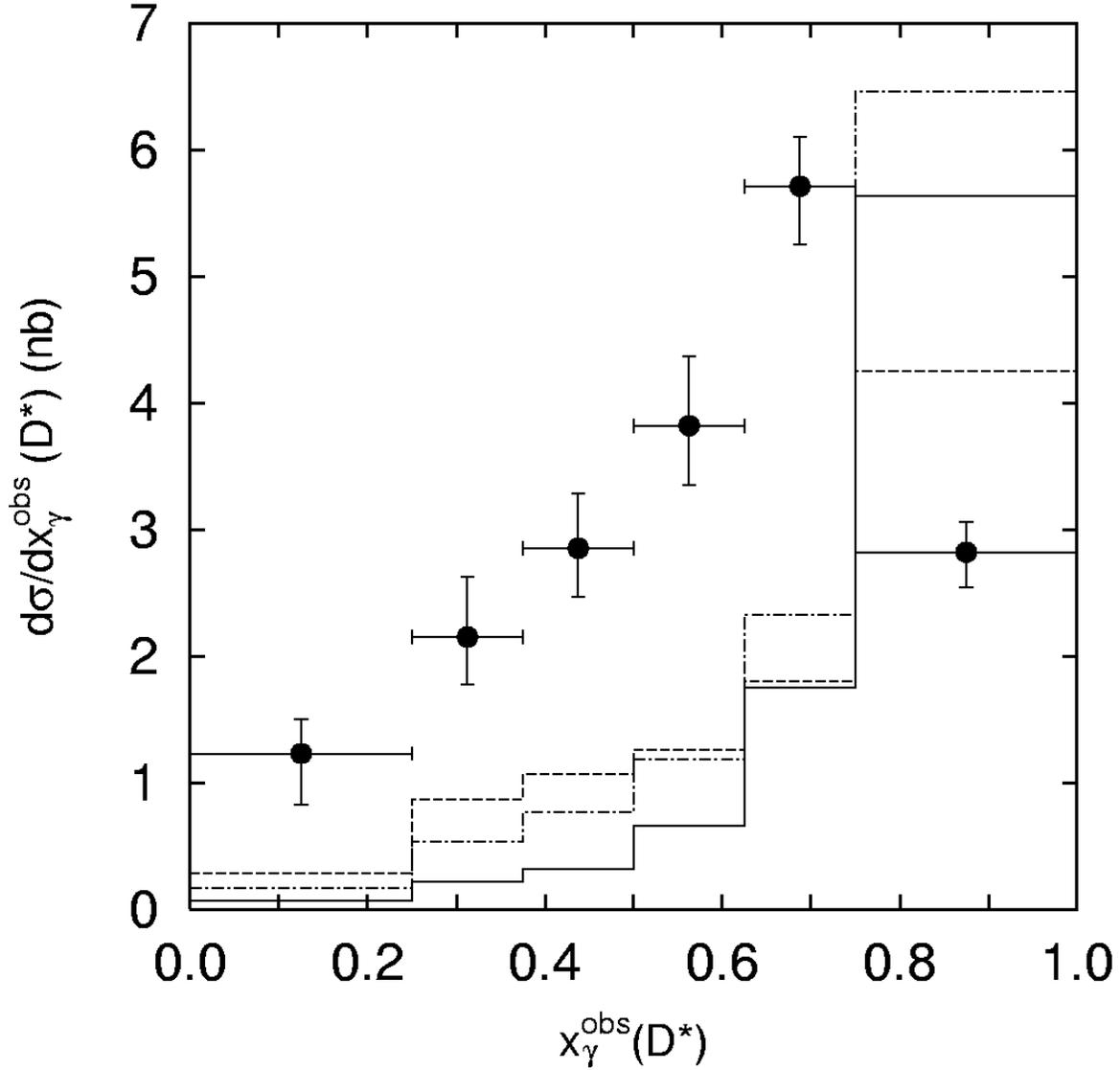, width = 22cm}
\caption{The $d\sigma/dx_\gamma^{\rm obs}(D^*)$ distribution of the 
associated $D^{*\pm}$ and jet production at HERA calculated in the 
kinematic range 
$Q^2 < 1$~GeV$^2$, $130 < W < 280$~GeV, 
$p_T > 3$~GeV, $|\eta| < 1.5$, $-1.5 < \eta^{\rm jet} < 2.4$
and $E_T^{\rm jet} > 6$~GeV. Notations of all histograms here are the 
same as in Fig.~1.
The experimental data are from ZEUS~[5].}
\label{fig7a}
\end{figure}

\newpage

\begin{figure}
\epsfig{figure=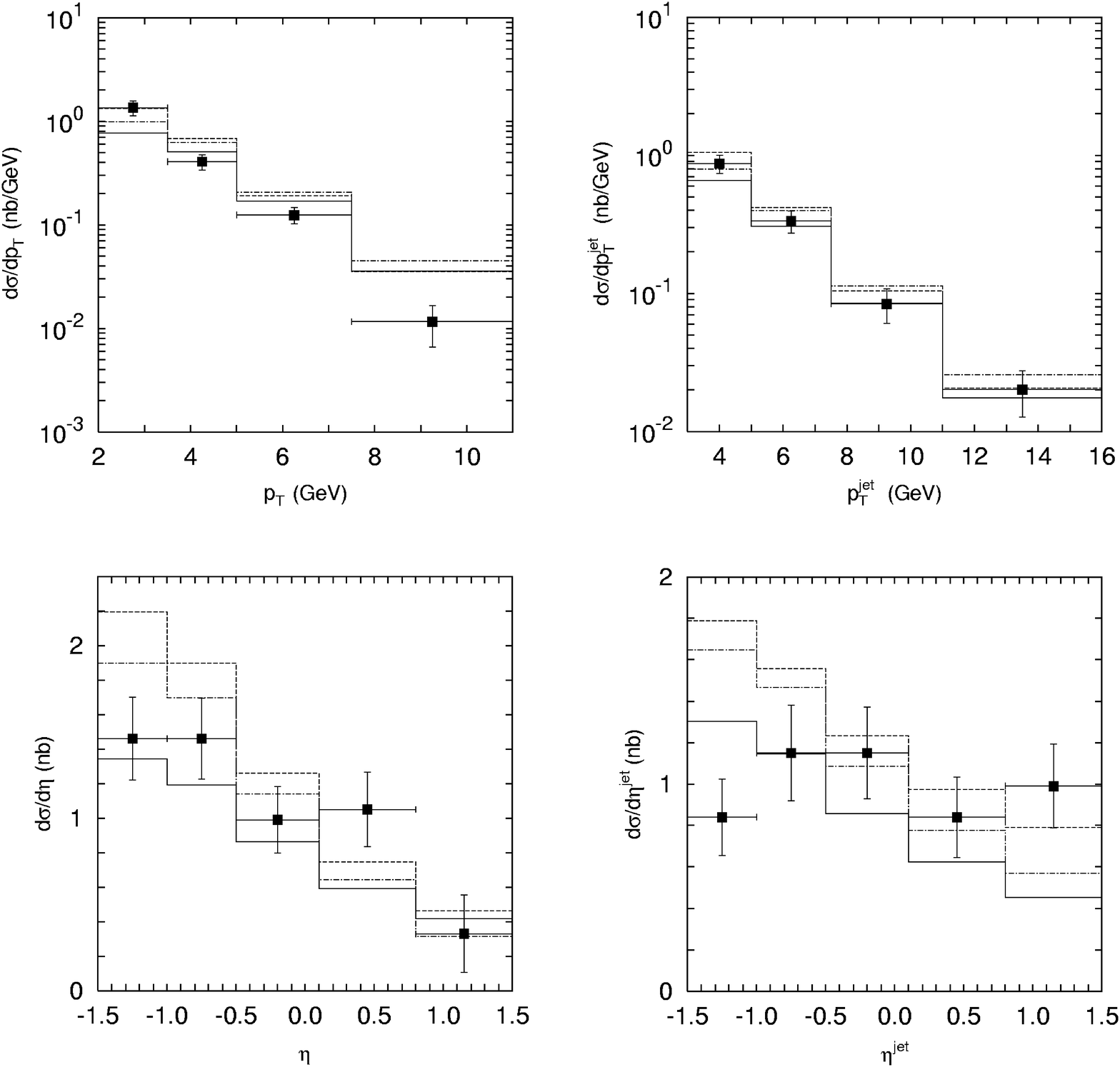, width = 17cm}
\caption{The differential cross sections
of the $D^{*\pm}$ and untagged jet production at 
HERA calculated in the kinematic range 
$Q^2 < 0.01$~GeV$^2$, $0.29 < y < 0.65$, $p_T > 2$~GeV,
$-1.5 < \eta < 1.5$, $p_T^{\rm jet} > 3$~GeV and
$-1.5 < \eta^{\rm jet} < 1.5$.
Notations of all histograms here are the 
same as in Fig.~1. The experimental data are from H1~[6].}
\label{fig6}
\end{figure}

\newpage

\begin{figure}
\epsfig{figure=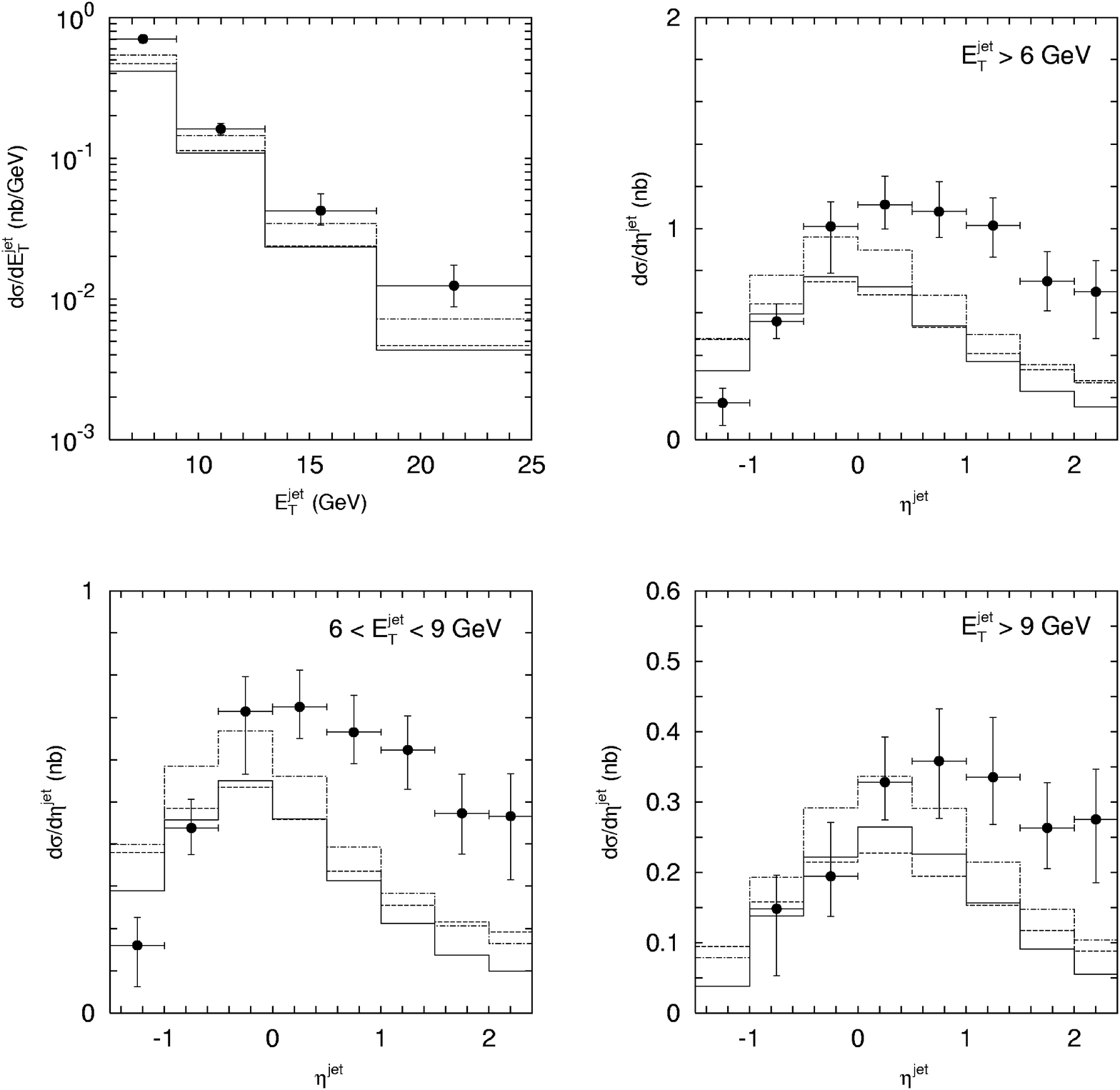, width = 17cm}
\caption{The differential cross sections
of the $D^{*\pm}$ and untagged jet production at 
HERA calculated in the kinematic range 
$Q^2 < 1$~GeV$^2$, $130 < W < 280$~GeV, 
$p_T > 3$~GeV, $|\eta| < 1.5$, $-1.5 < \eta^{\rm jet} < 2.4$
and $E_T^{\rm jet} > 6$~GeV. Notations of all histograms here are the 
same as in Fig.~1. The experimental data are from ZEUS~[5].}
\label{fig7}
\end{figure}

\newpage

\begin{figure}
\epsfig{figure=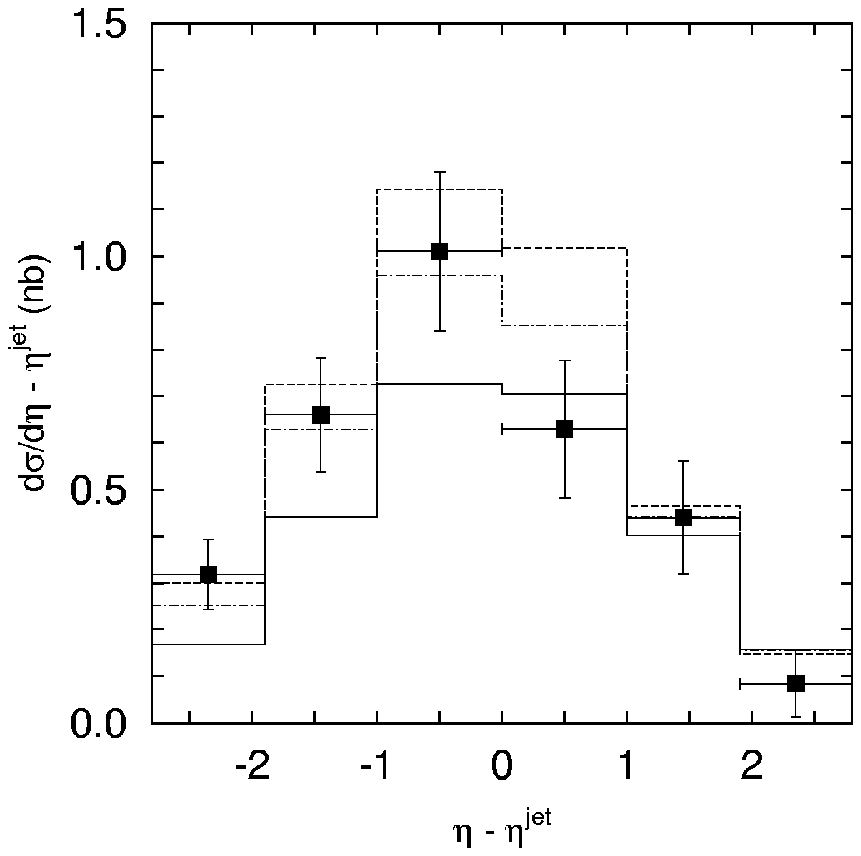, width = 22cm}
\caption{The cross section 
of the $D^{*\pm}$ and untagged jet production
as a function of $\eta - \eta^{\rm jet}$
calculated in the kinematic range 
$Q^2 < 0.01$~GeV$^2$, $0.29 < y < 0.65$, $p_T > 2$~GeV,
$-1.5 < \eta < 1.5$, $p_T^{\rm jet} > 3$~GeV and
$-1.5 < \eta^{\rm jet} < 1.5$. Notations of all histograms here are the 
same as in Fig.~1. The experimental data are from H1~[6].}
\label{fig8}
\end{figure}

\newpage

\begin{figure}
\epsfig{figure=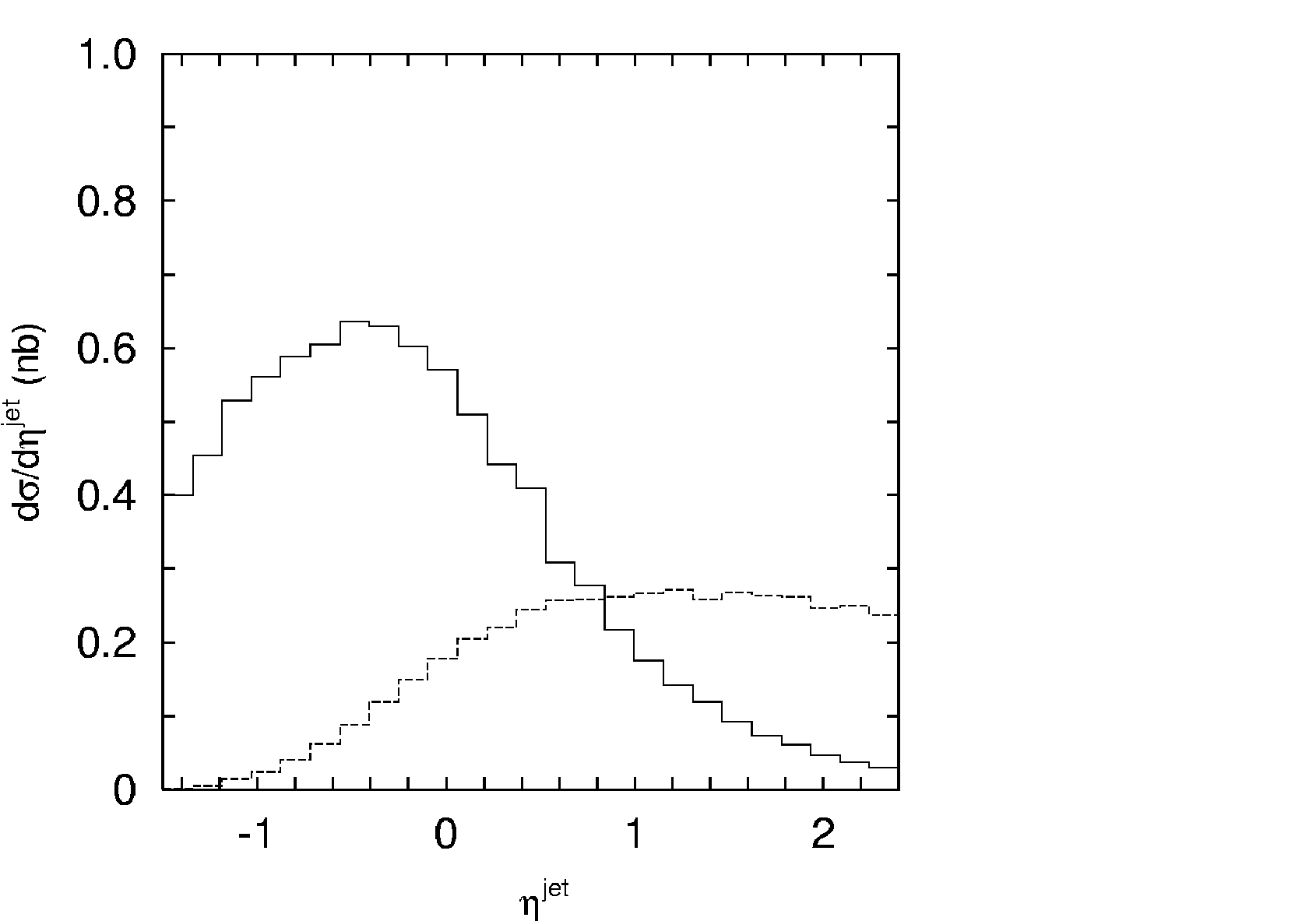, width = 22cm}
\caption{Direct (solid histogram) and resolved photon 
(dashed histogram) components of our central predictions
of the differential cross section $d\sigma/d\eta^{\rm jet}$
calculated in the kinematic range 
$Q^2 < 1$~GeV$^2$, $130 < W < 280$~GeV, 
$p_T > 3$~GeV, $|\eta| < 1.5$, $-1.5 < \eta^{\rm jet} < 2.4$
and $E_T^{\rm jet} > 6$~GeV. The KMR unintegrated
gluon densities in a proton and in a photon have been used.}
\label{fig9}
\end{figure}

\newpage

\begin{figure}
\epsfig{figure=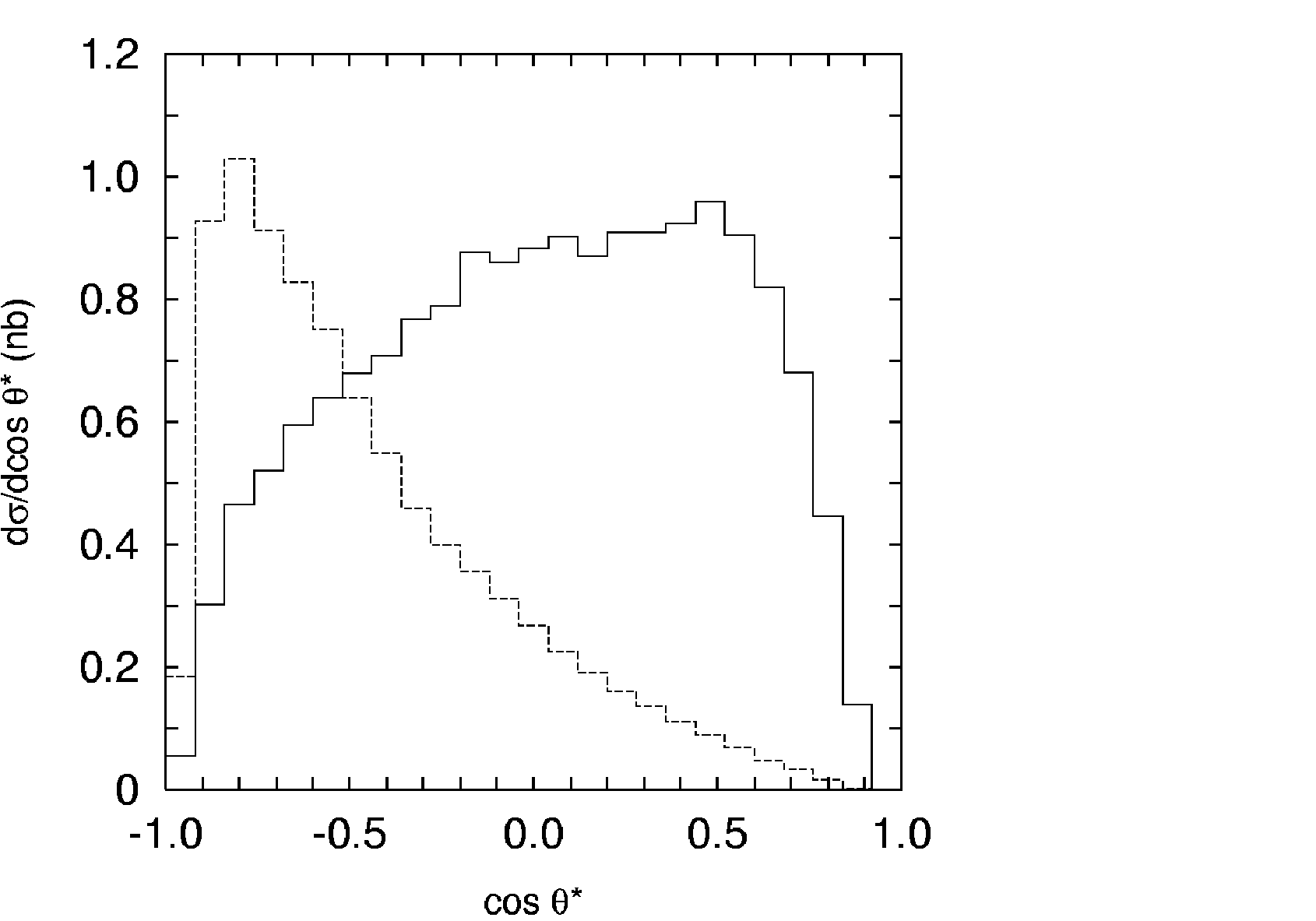, width = 22cm}
\caption{Direct (solid histogram) and resolved photon 
(dashed histogram) components of our central predictions
of the differential cross section $d\sigma/d\cos \theta^{*}$
calculated in the kinematic range 
$Q^2 < 1$~GeV$^2$, $130 < W < 280$~GeV, 
$p_T > 3$~GeV, $|\eta| < 1.5$, $-1.5 < \eta^{\rm jet} < 2.4$
and $E_T^{\rm jet} > 6$~GeV. The KMR unintegrated
gluon densities in a proton and in a photon have been used.}
\label{fig12}
\end{figure}

\newpage

\begin{figure}
\epsfig{figure=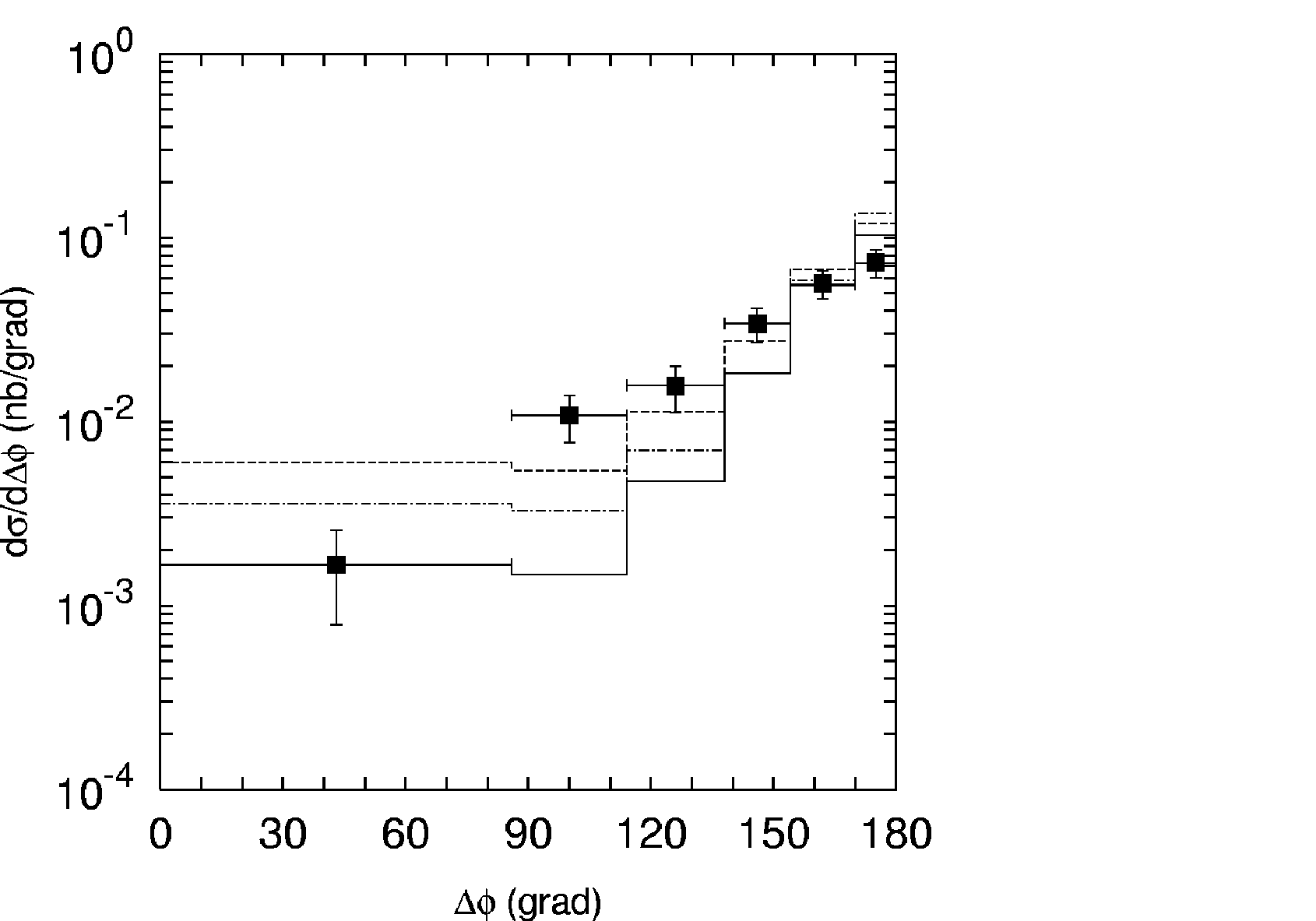, width = 22cm}
\caption{The cross section 
of the $D^{*\pm}$ and untagged jet production
as a function of $\Delta \phi$
calculated in the kinematic range 
$Q^2 < 0.01$~GeV$^2$, $0.29 < y < 0.65$, $p_T > 2$~GeV,
$-1.5 < \eta < 1.5$, $p_T^{\rm jet} > 3$~GeV and
$-1.5 < \eta^{\rm jet} < 1.5$. 
Notations of all histograms here are the 
same as in Fig.~1. The experimental data are from H1~[6].}
\label{fig13}
\end{figure}

\newpage

\begin{figure}
\epsfig{figure=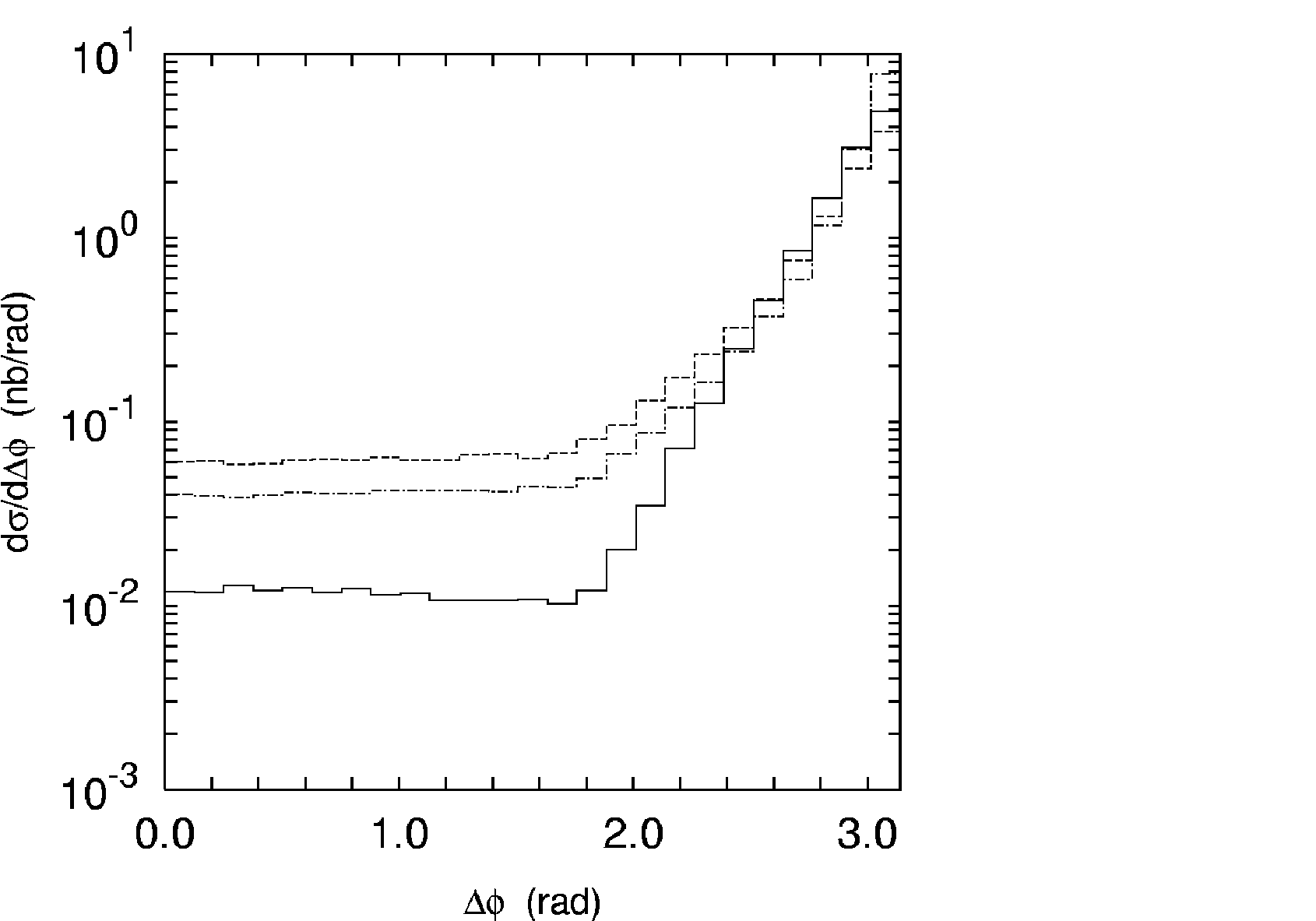, width = 22cm}
\caption{The cross section 
of the $D^{*\pm}$ and untagged jet production
as a function of $\Delta \phi$
calculated in the kinematic range 
$Q^2 < 1$~GeV$^2$, $130 < W < 280$~GeV, 
$p_T > 3$~GeV, $|\eta| < 1.5$, $-1.5 < \eta^{\rm jet} < 2.4$,
$E_T^{\rm jet} > 6$~GeV and $x_\gamma^{\rm obs}(D^{*}) > 0.75$.
Notations of all histograms here are the 
same as in Fig.~1.}
\label{fig14}
\end{figure}

\newpage

\begin{figure}
\epsfig{figure=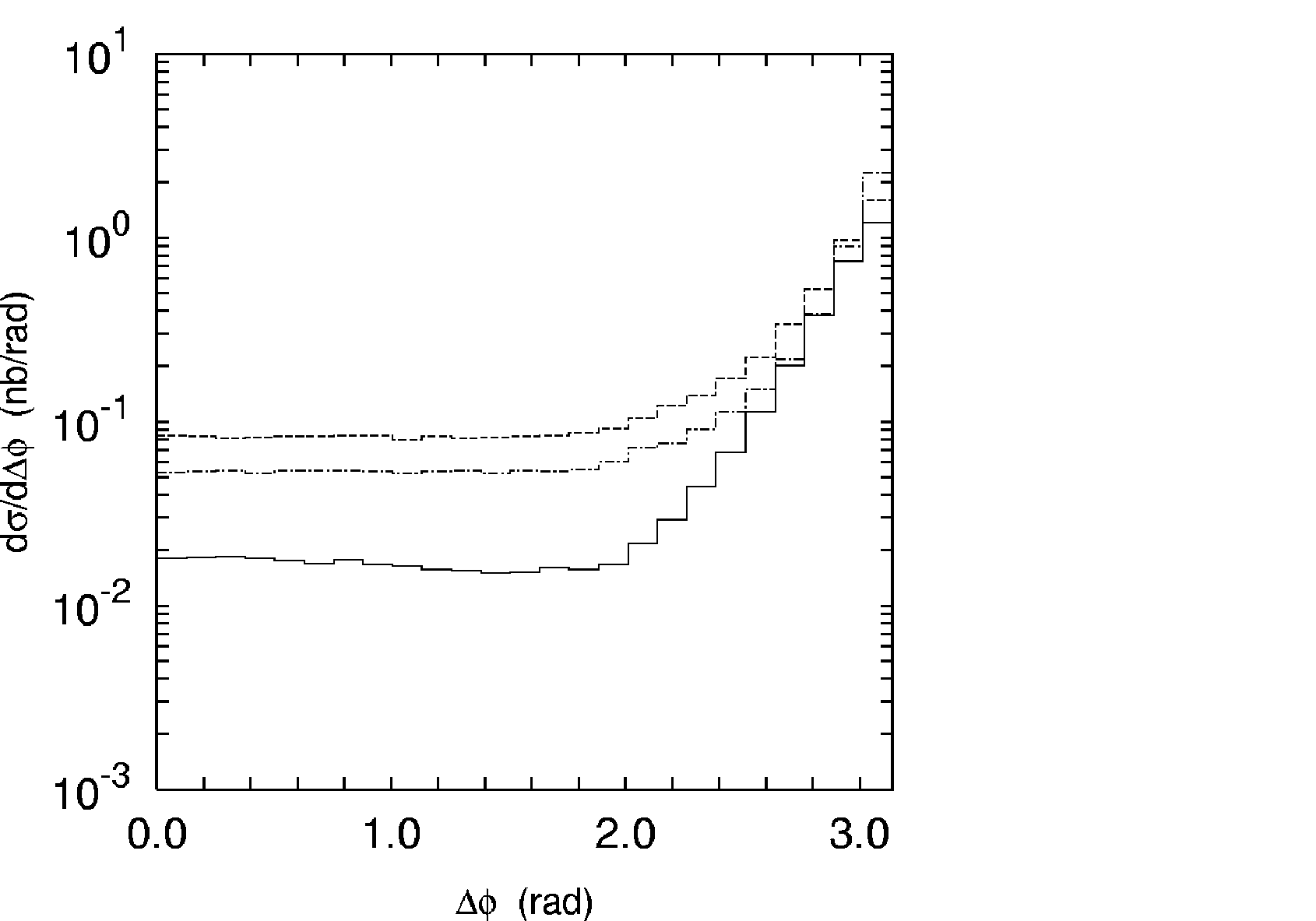, width = 22cm}
\caption{The cross section 
of the $D^{*\pm}$ and untagged jet production
as a function of $\Delta \phi$
calculated in the kinematic range 
$Q^2 < 1$~GeV$^2$, $130 < W < 280$~GeV, 
$p_T > 3$~GeV, $|\eta| < 1.5$, $-1.5 < \eta^{\rm jet} < 2.4$,
$E_T^{\rm jet} > 6$~GeV and $x_\gamma^{\rm obs}(D^{*}) < 0.75$.
Notations of all histograms here are the 
same as in Fig.~1.}
\label{fig15}
\end{figure}


\begin{thebibliography}{44}

\bibitem{1} C.~Adloff {\sl et al.} (H1 Collaboration), Nucl. Phys. B {\bf 545}, 21 (1999).
\bibitem{2} J.~Breitweg {\sl et al.} (ZEUS Collaboration), Eur. Phys. J. C {\bf 6}, 67 (1999).
\bibitem{3} S.~Chekanov {\sl et al.} (ZEUS Collaboration), Phys. Lett. B {\bf 565}, 87 (2003).
\bibitem{4} A.~Aktas {\sl et al.} (H1 Collaboration), Phys. Lett. B {\bf 621}, 56 (2005).
\bibitem{5} S.~Chekanov {\sl et al.} (ZEUS Collaboration), Nucl. Phys. B {\bf 729}, 492 (2005).
\bibitem{6} A.~Aktas {\sl et al.} (H1 Collaboration), DESY 06-110.
\bibitem{7} S.~Catani, M.~Ciafoloni and F.~Hautmann, Nucl. Phys. B {\bf 366}, 135 (1991).
\bibitem{8} J.C.~Collins and R.K.~Ellis, Nucl. Phys. B {\bf 360}, 3 (1991).
\bibitem{9} L.V.~Gribov, E.M.~Levin, and M.G.~Ryskin, Phys. Rep. {\bf 100}, 1 (1983).
\bibitem{10} E.M.~Levin, M.G.~Ryskin, Yu.M.~Shabelsky and A.G.~Shuvaev, Sov. J. Nucl. Phys. {\bf 53}, 657 (1991).
\bibitem{11} A.V.~Lipatov and N.P.~Zotov, Eur. Phys. J. C {\bf 47}, 643 (2006).
\bibitem{12} A.V.~Lipatov and N.P.~Zotov, Phys. Rev. D {\bf 73}, 114018 (2006);\\
  A.V.~Lipatov and N.P.~Zotov, JHEP {\bf 0608}, 043 (2006).
\bibitem{13} A.V.~Lipatov and N.P.~Zotov, Phys. Rev. D {\bf 72}, 054002 (2005).  
\bibitem{14} A.V.~Lipatov and N.P.~Zotov, DESY 05-157, to be published in J. Phys. G (2007).
\bibitem{15} S.P.~Baranov, N.P.~Zotov and A.V.~Lipatov, Phys. Atom. Nucl. {\bf 67}, 834 (2004).
\bibitem{16} E.A.~Kuraev, L.N.~Lipatov, and V.S.~Fadin, Sov. Phys. JETP {\bf 44}, 443 (1976);\\
  E.A.~Kuraev, L.N.~Lipatov, and V.S.~Fadin, Sov. Phys. JETP {\bf 45}, 199 (1977);\\
  I.I.~Balitsky and L.N.~Lipatov, Sov. J. Nucl. Phys. {\bf 28}, 822 (1978).
\bibitem{17} M.~Ciafaloni, Nucl. Phys. B {\bf 296}, 49 (1988);\\
  S.~Catani, F.~Fiorani, and G.~Marchesini, Phys. Lett. B {\bf 234}, 339 (1990);\\
  S.~Catani, F.~Fiorani, and G.~Marchesini, Nucl. Phys. B {\bf 336}, 18 (1990);\\
  G.~Marchesini, Nucl. Phys. B {\bf 445}, 49 (1995).
\bibitem{18} V.N.~Gribov and L.N.~Lipatov, Yad. Fiz. {\bf 15}, 781 (1972);\\
  L.N.~Lipatov, Sov. J. Nucl. Phys. {\bf 20}, 94 (1975);\\
  G.~Altarelly and G.~Parizi, Nucl. Phys. B {\bf 126}, 298 (1977);\\
  Y.L.~Dokshitzer, Sov. Phys. JETP {\bf 46}, 641 (1977).
\bibitem{19} S.~Frixione, P.~Nason and G.~Ridolfi, Nucl. Phys. B {\bf 454}, 3 (1995).
\bibitem{20} J.~Binnewies, B.A.~Kniehl and G.~Kramer, Z. Phys. C {\bf 76}, 677 (1997);\\ 
  B.A.~Kniehl, G.~Kramer and M.~Spira, Z. Phys. C {\bf 76}, 689 (1997);\\
  J.~Binnewies, B.A.~Kniehl and G.~Kramer, Phys. Rev. D {\bf 48}, 014014 (1998).
\bibitem{21} J.C.~Collins, Phys. Rev. D {\bf 58}, 094002 (1998);\\ 
  M.A.G.~Aivazis, J.C.~Collins, F.I.~Olness and W.-K.~Tung, Phys. Rev. D {\bf 50}, 3102 (1994);\\ 
  F.I.~Olness, R.J.~Scalise and W.-K.~Tung, Phys. Rev. D {\bf 59}, 014506 (1999);\\ 
  A.~Chuvakin, J.~Smith and W.L.~van~Neerven, Phys. Rev. D {\bf 61}, 096004 (2000). 
\bibitem{22} S.P.~Baranov and N.P.~Zotov, Phys. Lett. B {\bf 491}, 111 (2000).
\bibitem{23} S.P.~Baranov, H.~Jung, L.~J\"onsson, S.~Padhi and N.P.~Zotov, Eur. Phys. J. C {\bf 24}, 425 (2002).
\bibitem{24} M.~Luszczak and A.~Szczurek, Phys. Lett. B {\bf 594}, 291 (2004);\\
  Phys. Rev. D {\bf 73}, 054028 (2006).
\bibitem{25} H.~Jung, Comput. Phys. Comm. {\bf 143}, 100 (2002). 
\bibitem{26} A.V.~Lipatov and N.P.~Zotov, Eur. Phys. J. C {\bf 41}, 163 
(2005).
\bibitem{27} H.~Jung, Mod. Phys. Lett. A {\bf 19}, 1 (2004).
\bibitem{28} J.~Kwiecinski, A.D.~Martin and A.~Stasto, Phys. Rev. D {\bf 56}, 
3991 (1997).
\bibitem{29} M.A.~Kimber, A.D.~Martin and M.G.~Ryskin, Phys. Rev. D {\bf 63}, 
114027 (2001);\\
  G.~Watt, A.D.~Martin and M.G.~Ryskin, Eur. Phys. J. C {\bf 31}, 73 (2003).
\bibitem{30} G.P.~Lepage, J. Comput. Phys. {\bf 27}, 192 (1978).
\bibitem{31} B.~Andersson {\sl et al.} (Small-$x$ Collaboration), Eur. Phys. 
J. C {\bf 25}, 77 (2002).
\bibitem{32} J.~Andersen {\sl et al.} (Small-$x$ Collaboration), Eur. Phys. 
J. C {\bf 35}, 67 (2004).
\bibitem{33} J.~Andersen {\sl et al.} (Small-$x$ Collaboration), Eur. Phys. 
J. C {\bf 48}, 53 (2006).
\bibitem{34} J.~Kwiecinski, A.D.~Martin and A.~Sutton, Phys. Rev. D {\bf 52}, 1445 (1995);\\
  J.~Kwiecinski, A.D.~Martin and J. Outhwaite,  Eur. Phys. J. C {\bf 9}, 611 (2001).
\bibitem{35} M.~Gl\"uck, E.~Reya and A.~Vogt, Phys. Rev. D {\bf 46}, 1973 (1992);\\
  M.~Gl\"uck, E.~Reya and A.~Vogt, Z. Phys. C {\bf 67}, 433 (1995). 
\bibitem{36} S.~Chekanov {\sl et al.} (ZEUS Collaboration), DESY 06-125.
\bibitem{37} M.A.~Kimber, A.D.~Martin and M.G.~Ryskin, Eur. Phys. J. C {\bf 12}, 655 (2001).
\bibitem{38} M.~Hansson, H.~Jung and L.~J\"onsson, hep-ph/0402019.
\bibitem{39} L.~Motyka and N.~Timneanu, Eur. Phys. J. C {\bf 27}, 73 (2003).
\bibitem{40} C.~Peterson, D.~Schlatter, I.~Schmitt, and P.~Zerwas, Phys. Rev. D {\bf 27}, 105 (1983).
\bibitem{41} P.~Nason and C.~Oleari, Nucl. Phys. B {\bf 565}, 245 (2000).
\bibitem{42} K.~Ackerstaff {\sl et al.} (OPAL Collaboration), Eur. Phys. J. C {\bf 1}, 439 (1997).
\bibitem{43} S.~Frixione, M.L.~Mangano, P.~Nason and G.~Ridolfi, Phys. Lett. B {\bf 348}, 633 (1995).
\bibitem{44} T.~Sj\"ostrand {\sl et al.}, Comput. Phys. Comm. {\bf 135}, 238 (2001).
\bibitem{45} G.~Heinrich and B.A.~Kniehl, Phys. Rev. D {\bf 70}, 094035 (2004).

\end{thebibliography}
\end{document}